\documentclass[aps,physrev,10pt,floatfix,reprint]{revtex4-2}

\pdfoutput=1
\usepackage{graphicx}
\usepackage{dcolumn}
\usepackage{bm}
\usepackage{hyperref}
\usepackage[mathlines]{lineno}

\usepackage[
scale=0.7, marginratio={1:1, 2:3}, ignoreall,
text={7in,10in},centering,
]{geometry}
\usepackage[utf8]{inputenc}
\usepackage{amsmath}
\usepackage{amstext}
\usepackage{graphicx}
\usepackage{esint}
\usepackage{geometry}
\usepackage{hyperref}
\usepackage{amsfonts}
\usepackage{nicefrac}

\hypersetup{
    colorlinks=true,
    linkcolor=blue,
    filecolor=magenta,      
    urlcolor=cyan,
    pdftitle={Overleaf Example},
    pdfpagemode=FullScreen,
    }
\usepackage{changes}

\begin{document}

\title{The autoregressive neural network architecture of the Boltzmann distribution of pairwise interacting spins systems}
\author{Indaco, Biazzo}
\email{indaco.biazzo@epfl.ch}
\affiliation{
Ecole Polytechnique Fédérale de Lausanne (EPFL). Statistical Physics of Computation (SPOC) lab.
}

\begin{abstract}
    Generative Autoregressive Neural Networks (ARNNs) have recently demonstrated exceptional results in image and language generation tasks, contributing to the growing popularity of generative models in both scientific and commercial applications. This work presents an exact mapping of the Boltzmann distribution of binary pairwise interacting systems into autoregressive form. The resulting ARNN architecture has weights and biases of its first layer corresponding to the Hamiltonian's couplings and external fields, featuring widely used structures such as the residual connections and a recurrent architecture with clear physical meanings. Moreover, its architecture's explicit formulation enables the use of statistical physics techniques to derive new ARNNs for specific systems. As examples, new effective ARNN architectures are derived from two well-known mean-field systems, the Curie-Weiss and Sherrington-Kirkpatrick models, showing superior performance in approximating the Boltzmann distributions of the corresponding physics model compared to other commonly used architectures. The connection established between the physics of the system and the neural network architecture provides a means to derive new architectures for different interacting systems and interpret existing ones from a physical perspective.
\end{abstract}
    
    
\maketitle
\section{Introduction} 
The cross-fertilization between machine learning and statistical physics, in particular of disordered systems, has a long history \cite{doi:10.1073/pnas.79.8.2554, PhysRevA.32.1007}.
Recently, the development of deep neural network frameworks \cite{bengioNatureDeepLearning2015} have been applied to statistical physics problems \cite{RevModPhys.91.045002} spanning a wide range of domains, including quantum mechanics \cite{doi:10.1126/science.aag2302, Nieuwenburg2017}, 
classical statistical physics \cite{Carrasquilla2017, Wu2019}, chemical and biological physics \cite{noe2019boltzmann,jumper2021highly}.
On the other hand, techniques borrowed from statistical physics have been used to shed light on the behavior of Machine Learning algorithms \cite{doi:10.1080/00018732.2016.1211393, Nguyen2017}, and even to suggest training or architecture frameworks \cite{Chaudhari_2019, pmlr-v37-sohl-dickstein15}.
In recent years, the introduction of deep generative autoregressive models \cite{pmlr-v37-germain15, NIPS2016_b1301141}, like transformers \cite{NIPS2017_3f5ee243}, has been a breakthrough in the field, generating images and text with a quality comparable to human-generated ones \cite{https://doi.org/10.48550/arxiv.2005.14165}.  
The introduction of deep ARNNs was motivated as a flexible and general approach to sampling from a probability distribution learned from data \cite{pmlr-v32-gregor14, pmlr-v15-larochelle11a, pmlr-v48-oord16}. 

In classical statistical physics, the ARNN was introduced, in a variational setting, to sample from a Boltzmann distribution (or equivalently an energy-based model \cite{pmlr-v97-durkan19a}) as an improvement over the standard variational approach relying on the high expressiveness of the ARNNs \cite{Wu2019}. 

Then similar approaches have been used in different contexts, and domains of classical \cite{10.1103/physreve.101.023304,PhysRevE.101.053312,PhysRevE.103.012103,PhysRevResearch.3.L042024,10.1038/s42256-021-00401-3} and quantum statistical physics \cite{10.1103/physrevlett.128.090501,PhysRevA.102.062413,PhysRevLett.124.020503,PhysRevResearch.2.023358, Liu_2021, Barrett2022, Cha_2022}. The ability of ARNNs to efficiently generate samples, thanks to the ancestral sampling procedure, opened the way to overcome the slowdown of Monte-Carlo methods for frustrated or complex systems, although two recent works questioned the real gain in very frustrated systems \cite{condmat7020038,https://doi.org/10.48550/arxiv.2210.11145}. 

The use of ARNNs in statistical physics problems has largely relied on pre-existing neural network architectures which may not be well-suited for the particular problem at hand. This approach has been largely favored due to the high expressive capacity of ARNNs, which can encapsulate the complexity of the Boltzmann probability distribution, remapped in an autoregressive form, within their parameters that, typically, grow polynomially with system size. To encode this complexity exactly, however, one might expect the need for an exponentially large number of parameters.
This work aims to demonstrate how knowledge of the physics model can inform the design of more effective ARNN architectures. I will present the derivation of an ARNN architecture that encodes exactly the classical Boltzmann distribution associated with a general pairwise interacting Hamiltonian of binary variables. Despite the generality of the Hamiltonian, the resulting architecture exhibits interesting properties: the first layer's parameters, which scale polynomially with the system size, are fixed by the Hamiltonian parameters. Additionally, the derivation introduces both residual connections and recurrent structures, each with a clear physical interpretation.\\
As expected for the exact architecture of the general case, the resulting deep ARNN architecture has the number of hidden layer parameters scaling exponentially with the system's size. In the general case, it is possible to approximate these hidden layers with feed-forward neural network structures containing a polynomial number of free parameters. The advantage of this approach over existing architectures is that the first layer's parameters can be fixed by the Hamiltonian parameters, reducing the number of parameters to be learned and trained.
For instance the proposed architecture could be used in accelerating Markov chain simulations \cite{10.1103/physreve.101.023304,PhysRevE.101.053312}. 
The quality of the approximation of the Boltzmann distribution relies on both the architecture of the feed-forward neural network used and the complexity of the problem being tackled.
However, 
the clear physical interpretation of the architecture allows us to leverage problem-specific knowledge to develop specific feed-forward neural network architectures. As an example, standard statistical physics techniques will be used in the following to find feasible ARNN architecture for specific Hamiltonian. To showcase the potential of the derived representation, the ARNN architectures for two well-known mean-field models are derived: the Curie-Weiss (CW) and the Sherrington-Kirkpatrick model (SK). These fully connected models are chosen due to their paradigmatic role in the history of statistical physics systems. 

The CW model, despite its straightforward Hamiltonian, was one of the first models explaining the behavior of ferromagnet systems, displaying a second-order phase transition \cite{kadanoff2000statistical}. In this case, an exact ARNN architecture at finite N and an approximated architecture in the thermodynamic limit are derived, both with a number of parameters scaling polynomially with the system's size. 

The SK model \cite{PhysRevLett.35.1792} is a fully connected spin glass model of disordered magnetic materials. The system admits an analytical solution in the thermodynamic limit, Parisi's celebrated \cite{Nobel2021} k-step replica symmetric breaking (k-RSB) solution \cite{PARISI1979203, PhysRevLett.43.1754}. The complex many-valley landscape of the Boltzmann probability distribution captured by the k-RSB solution of the SK model is the key concept that unifies the description of many different problems, and similar replica computations are applied to very different domains like neural networks \cite{Gardner_1987, PhysRevLett.55.1530}, optimizations \cite{doi:10.1126/science.1073287}, inference problems \cite{doi:10.1080/00018732.2016.1211393}, or in characterizing the jamming of hard spheres \cite{RevModPhys.82.789, PhysRevLett.102.195701}. In this work, an ARNN architecture of the Boltzmann distribution of the SK model for a single instance of disorder, with a finite number of variables will be shown. The derivation will be based on the k-RSB solution, resulting in a deep ARNN architecture with parameters scaling polynomially with the system size. 

In the following, I will first present the ARNN architecture for the Boltzmann distribution of the pairwise interacting systems, and then demonstrate how to derive the new architecture for the CW and SK models. Finally, the last section compares the performance of the derived ARNN architectures with standard ARNN architectures used in the literature.
            
\section{Autoregressive architecture of the Boltzmann distribution of pairwise interacting systems \label{sec:ARNN_boltzmann}}

The Boltzmann probability distribution of a given Hamiltonian $H[\mathbf{x}]$ of a set of $N$ binary spin variables $\mathbf{x}=(x_1, x_2,...x_N)$ at inverse temperature $\beta$ is $P_{B}(\mathbf{x}) = \nicefrac{e^{-\beta H\left(\mathbf{x}\right)}}{Z}$, where $Z=\sum_{\mathbf{x}}e^{-\beta H\left(\mathbf{x}\right)}$ is the normalization factor.
It is generally challenging to compute marginals and average quantities when $N$ is large and in particular, generate samples on frustrated systems. By defining the sets of variables $\mathbf{x}_{<i}=\left(x_{1},x_{2}\dots x_{i-1}\right)$ and $\mathbf{x}_{>i}=\left(x_{i+1},x_{i+2}\dots x_{N}\right)$ respectively with an index smaller and larger than $i$, then if we can rewrite the Boltzmann distribution in the autoregressive form: $P_{B}\left(\mathbf{x}\right)=\prod_{i}P\left(x_{i}|\mathbf{x}_{<i}\right)$, it becomes straightforward to produce independent samples from it, thanks to the ancestral sampling procedure \cite{Wu2019}. It has been proposed \cite{Wu2019} to use a variational approach to approximate the Boltzmann distribution with trial autoregressive probability distributions where each conditional probability is represented by a feed-forward neural network with a set of parameters ${\theta}$,
$Q^{\theta}\left(\mathbf{x}\right)=\prod_{i}Q^{\theta_i}\left(x_{i}|\mathbf{x}_{<i}\right)$.

The parameters ${\theta}$ can be learned minimizing  the variational free energy of the system:

\begin{equation}
    F[P]= \sum_{\left\{ \mathbf{x} \right\}}P[\mathbf{x}]\left[\frac{1}{\beta}\log P[\mathbf{x}] + H[\mathbf{x}] \right].
    \label{eq:kl}    
\end{equation}
Minimizing the variational free energy $F[Q^{\theta}]$ with respect to the parameters of the ARNN is equivalent to minimizing Kullback-Leibler divergence with the Boltzmann distribution as the target \cite{Wu2019}. The computation of $F[Q^{\theta}]$ and their derivatives with respect to the ARNN's parameters involve a summation over all the configurations of the systems, that grows exponentially with the system's size, making it unfeasible after a few spins. In practice, they are estimated summing over a subset of configurations sampled directly from the ARNN thanks to the ancestral sampling procedure\cite{Wu2019}. 
Besides the minimization procedure, the choice of the architecture of the neural networks is crucial to obtain a good approximation of the Boltzmann distribution. In the following, I will derive an exact ARNN architecture of the Boltzmann distribution of pairwise-interacting spins.

\subsection{The single variable conditional probability}

In the parametrization $Q^{\theta_i}\left(x_{i}=1|\mathbf{x}_{<i}\right)$ of the single variable conditional probability distribution $P\left(x_{i}=1|\mathbf{x}_{<i}\right)$ as a feed-forward neural network, the set of variables $\mathbf{x}_{<i}$ is the input, and a nested set of linear transformations, and non-linear activation functions is applied on them.

Usually, the last layer is a sigma function $\sigma(x)=\frac{1}{1+e^{-x}}$, assuring the output is between $0$ and $1$. The probability $Q^{\theta_i}\left(x_{i}=-1|\mathbf{x}_{<i}\right) = 1 - Q^{\theta_i}\left(x_{i}=1|\mathbf{x}_{<i}\right)$ is straightforward to obtain. The set of parameters $\theta_i$ are the weights and biases of the linear transformations.
In the following, I will rewrite the single variable conditional probability of the Boltzmann distribution as a feed-forward neural network.

The generic $i$-th conditional probability factor of the Boltzmann distribution can be rewritten in this form: 

\begin{equation}
    \label{eq:chain}
    \begin{split}
    & P\left(x_{i}|\mathbf{x}_{<i}\right)  = 
    \frac{P\left(\mathbf{x}_{<i+1}\right)}{P\left(\mathbf{x}_{<i}\right)}  = 
    \frac{\sum_{\mathbf{x}_{>i}}P\left(\mathbf{x}\right)}{\sum_{\mathbf{x}_{>i-1}}P\left(\mathbf{x}\right)} \\
    &=\frac{\sum_{\mathbf{x}_{>i}}e^{-\beta H}}{\sum_{\mathbf{x}_{>i-1}}e^{-\beta H}}  = 
    \frac{f\left(x_{i},\mathbf{x}_{<i}\right)}{\sum_{x_{i}}f\left(x_{i},\mathbf{x}_{<i}\right)}.
    \end{split}
\end{equation}
where I defined: 

\begin{equation}
f\left(x_{i}=\pm 1,\mathbf{x}_{<i}\right) = \sum_{\mathbf{x}_{>i}}e^{-\beta H}\delta_{x_i, \pm1}.  
\end{equation}
The $\delta_{a,b}$ is the Kronecker delta function that is one when the two values $(a,b)$ coincide and zero otherwise. Now imposing as last activation function a sigma function, with simple algebraic manipulations, we obtain: 

\begin{equation}
    \label{eq:sigma_log}
    \begin{split}
    & P\left(x_{i}=1|\mathbf{x}_{<i}\right) = 
    \frac{f\left(1,\mathbf{x}_{<i}\right)}{ f\left(1,\mathbf{x}_{<i}\right) + f\left(-1,\mathbf{x}_{<i}\right)} \\
    &=  \frac{1}{ 1 + \frac{f\left(-1,\mathbf{x}_{<i}\right)}{f\left(1,\mathbf{x}_{<i}\right)}}  = \sigma\left(\log\left[f\left(1,\mathbf{x}_{<i}\right)\right]-\log\left[f\left(-1,\mathbf{x}_{<i}\right)\right]\right)
    \end{split}
\end{equation}
Consider a generic two-body interaction Hamiltonian of binary spin variables $x_i \in \{-1,1\}$, $H = -\sum_{i<j} J_{ij} x_i x_j - \sum_{i} h_i x_i$, where the sets of $J_{ij}$ are the interaction couplings and $h_i$ are the external fields. Taking into account a generic variable $x_i$ the elements of the Hamiltonian can be grouped into the following five sets:

\begin{align*}
    H_{ss} &= -\sum_{s,s'<i}J_{ss'} x_s x_{s'} - \sum_{s<i} h_s x_s \\
    H_{si}[x_i=\pm 1] & =  \mp H_{si} = \mp (\sum_{s<i} J_{si} x_s + h_i)  \\
    H_{il}[x_i = \pm 1] & = \mp H_{il} = \mp \sum_{l>i} J_{il} x_l\\
    H_{sl} &= -\sum_{s<i,l>i}J_{sl} x_s x_{l}\\
    H_{ll} &= -\sum_{l,l'>i}J_{ll'} x_l x_{l'} - \sum_{l>i} h_l x_l \\
\end{align*}
where the dependence on the variable $x_i$ has been made explicit. Substituting these expressions in Eq.~(\ref{eq:sigma_log}), we obtain:

\begin{equation}
    \label{eq:conditional_ghann}
    \begin{split}
     P\left(x_{i}=1|\mathbf{x}_{<i}\right) = \sigma\bigg( 2 \beta H_{si}[\mathbf{x}_{<i}] +\log(\hat{\rho}_i^+[\mathbf{x}_{<i}]) \\
     - \log(\hat{\rho}_i^-[\mathbf{x}_{<i}])
    \bigg),   
    \end{split}
\end{equation}

where:

\begin{align}
    \hat{\rho}_i^{\pm} [\mathbf{x}_{<i}]  =& \sum_{\mathbf{x}_{>i}}  e^{-\beta(\pm H_{il} + H_{sl}[\mathbf{x}_{<i}] + H_{ll})} 
\label{eq:rho_ghann_tilde}
\end{align}

The set of elements $H_{ss}$ cancels out.

\begin{figure}[!ht]
    \includegraphics[width=0.45\textwidth]{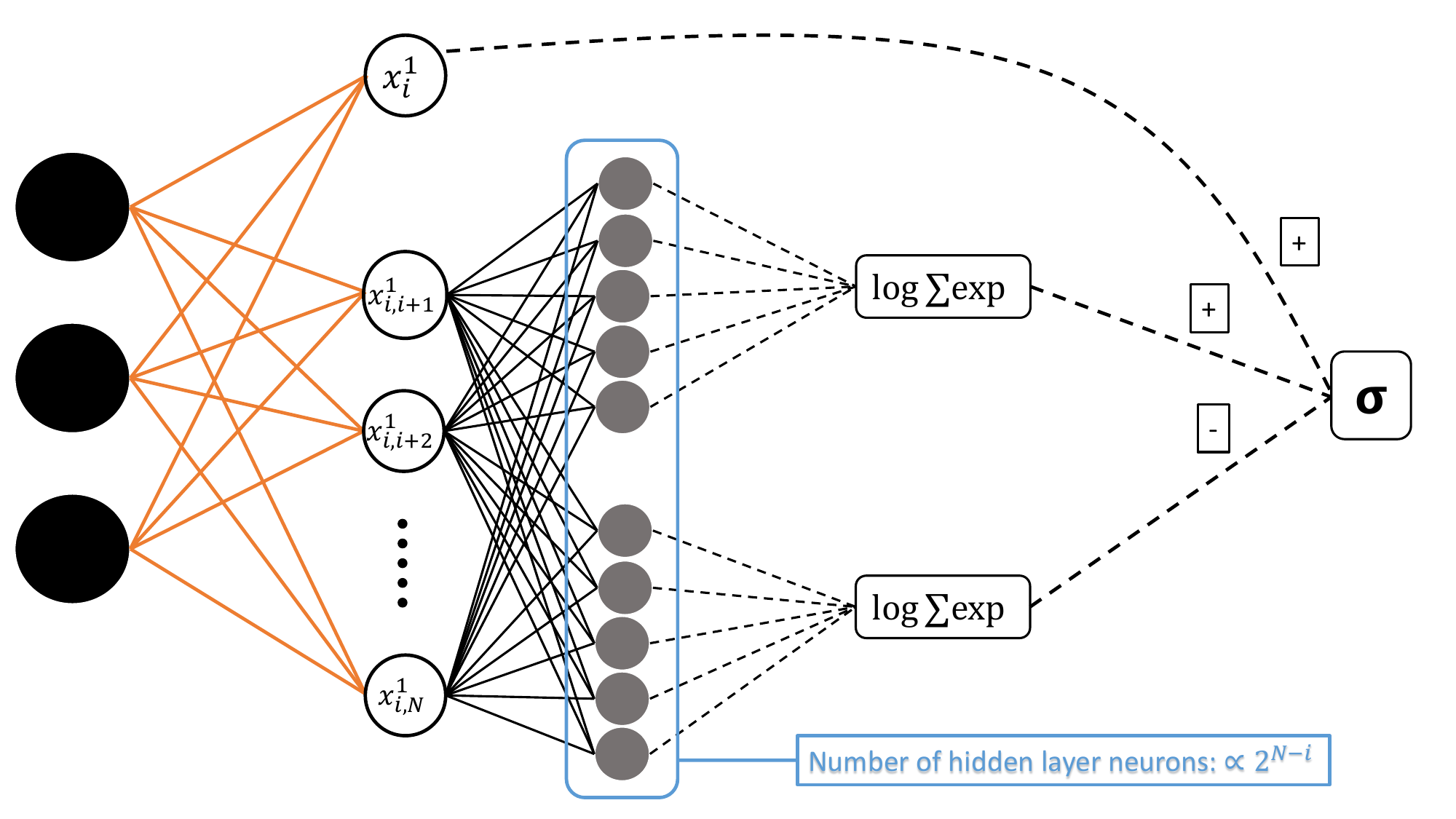}
    \caption{ \textbf{H$_2$ARNN} Architectures of a single Boltzmann conditional probability of a pairwise interacting Hamiltonian, Eq.~(\ref{eq:H2ANN}.) The $x_{<i}$ variables are the input, the output provides an estimation of the conditional probability $P (x_i=1 | \mathbf{x_{<i}})$. The first layer computes the $x^1_i$ and $x^1_{il}$ variables, see Eq.~(\ref{eq:x_i_first}), where the weight and bias, directly related to the Hamiltonian parameters, are shown in orange. The non-linear operators are represented by square symbols. The width of the second layer increases exponentially with the system size. The $\log\sum\exp(\mathbf{x})=\log \sum_i e^{x_i}$ represents the set of linear transformations and non-linear activation functions acting on the second layer.
    The last layer is the sigma function.}
    \label{fig:arch}
\end{figure}

The conditional probability, Eq.~(\ref{eq:conditional_ghann}), can be interpreted as a feed-forward neural network, following, starting from the input, the operation done on the variables $\mathbf{x}_{<i}$.
The first operation on the input is a linear transformation. Defining: 

\begin{align}
    \label{eq:x_i_first}
    x_i^1 &= 2 \beta H_{si} =  2 \beta( \sum_{s=1}^{i-1} J_{si} x_s + h_i),\\
    \label{eq:x_il_first}
    x_{il}^1 &= \sum_{s=1}^{i-1} J_{sl} x_s,
\end{align}

as outputs of the first layer (see Fig.~\ref{fig:arch}), we can write the conditional probability as a feed-forward neural network:

\begin{align}
    \label{eq:H2ANN}
        P_i\left(x_i=1 | \mathbf{x}_{<i}\right) & = 
     \sigma \left( x_i^1 + \log\rho_i^+ -\log\rho_i^- \right) \\
     \rho_i^{\pm} & = \sum_{c} e^{b_c^{\pm} + \sum_{l=i+1}^{N} w_{cl} x_{il}^1} \label{eq:rho_ghann}
\end{align}

As shown in Fig.~\ref{fig:arch}, a second linear transformation acts on the set of $x_{il}^1$ variables. 
The parameters of the second layer are

\begin{align}
    b_c^{\pm} &= \beta\sum_{l=i+1}^N (\pm J_{il} + h_l + \sum_{l'=l+1}^N J_{ll'}x^c_{l'}) x^c_l \\
    w_{cl} &=\beta x^c_l,
\end{align}
 where $c$ is the index of the configuration of the set of $\mathbf{x}_{>i}$ variables. This second linear transformation compute the $2^{N-i}$ possible values of the exponent in the $\rho_i^{\pm}$ functions, Eq.~(\ref{eq:rho_ghann}).
Next, the two functions $\rho_i^{\pm}$ are obtained by first applying the exponential function to the output of the second layer. Then, for each of $\rho_i^{\pm}$, we sum their elements and finally apply the logarithmic function.
As the last layer, the values $\log\rho_i^{\pm}$ and $x_i^1$ are combined with the right signs, and the sigma function is applied. The entire ARNN architecture of the Boltzmann distribution of the general pairwise interacting Hamiltonian ($\text{H}_2\text{ARNN}$) is depicted in Fig.~\ref{fig:arch}. The total number of parameters scales exponentially with the system size, making its direct application infeasible for the sampling process.
Nevertheless, the $\text{H}_2\text{ARNN}$ architecture shows some interesting features:
\begin{itemize}
    \item 
    The weights and biases of the first layer are the parameters of the Hamiltonian of the Boltzmann distribution. 
    As far as the author knows, this is the first time that this connection is derived.
    \item Residual connections among layers, due to the $x_i^1$ variables, naturally emerge from the derivation. 
    The importance of residual connections has recently been highlighted  \cite{10.48550/arxiv.1512.03385} and has become a crucial element in the success of the ResNet and transformer architectures \cite{vaswani2017attention}, in classification and generation tasks. They were presented as a way to improve the training of deep neural networks avoiding the exploding and vanishing gradient problem. In this context, they represent the direct interactions among the variable $x_i$ and all the previous variables $\mathbf{x}_{<i}$.

    \item The $\text{H}_2\text{ARNN}$ exhibits a recurrent structure  \cite{bengioNatureDeepLearning2015, https://doi.org/10.48550/arxiv.1506.00019}. 
    The first layer, as seen in Figure \ref{fig:arch}, is composed of a set of linear transformations (see eq.\ref{eq:x_i_first} and \ref{eq:x_il_first}). The set of $x^1_{il}=\sum_{s=1}^{i-1} J_{si} x_s$ variables, can be rewritten in recursive form observing that:

    \begin{equation}
        x^1_{il} = x^1_{i-1,l} + J_{i-1,l} x_{i-1}
    \end{equation}

    The output of the first layer of the conditional probability of the variable $i$ depend on the output of the first layer, $x^1_{i-1,l}$, of the previous $i-1$ conditional probability. In practice, we can explicitly write the following dependence: $P(x_i=1 | \mathbf{x}_{<i}) = P(x_i=1 | \mathbf{x}_{<i-1}$, $x^1_{i-1,l})$. These reduce the number of parameters of the first layers and could reduce its total computational cost if efficiently implemented.    
\end{itemize}

The most computationally demanding part of the H$_2$ARNN architecture is the computation of the $\rho_i^{\pm}$ functions, Eq.~(\ref{eq:rho_ghann}); their parameters scale exponentially with the system size, proportionally to $2^{N-i}$. However, generally, the $\rho_i^{\pm}$ functions can be approximated using standard feed-forward neural network structures, possessing a polynomial number of parameters. Here, the input variables are those of the first layer $\mathbf{x^1_{il}}$, while the parameters of the first layer remain unchanged, maintaining the skip connection. Instead of exploring this possibility, in the following, I will show how to derive new ARNN architectures for specific systems, based on statistical physics techniques.
In fact, the $\rho_i^{\pm}$ function can be interpreted as the partition function of a system, where the variables are the $\mathbf{x}_{>i}$ and the external fields are determined by the values of the variables $\mathbf{x}_{<i}$.
Based on this observation, in the following, I will show how to use standard tools of statistical physics to derive deep ARNN architectures that eliminate the exponential growth of the number of parameters. 

\section{Models}
\subsection{The Curie-Weiss (CW) model}

\begin{figure}[!h]
    \centering 
    \includegraphics[width=0.48\textwidth]{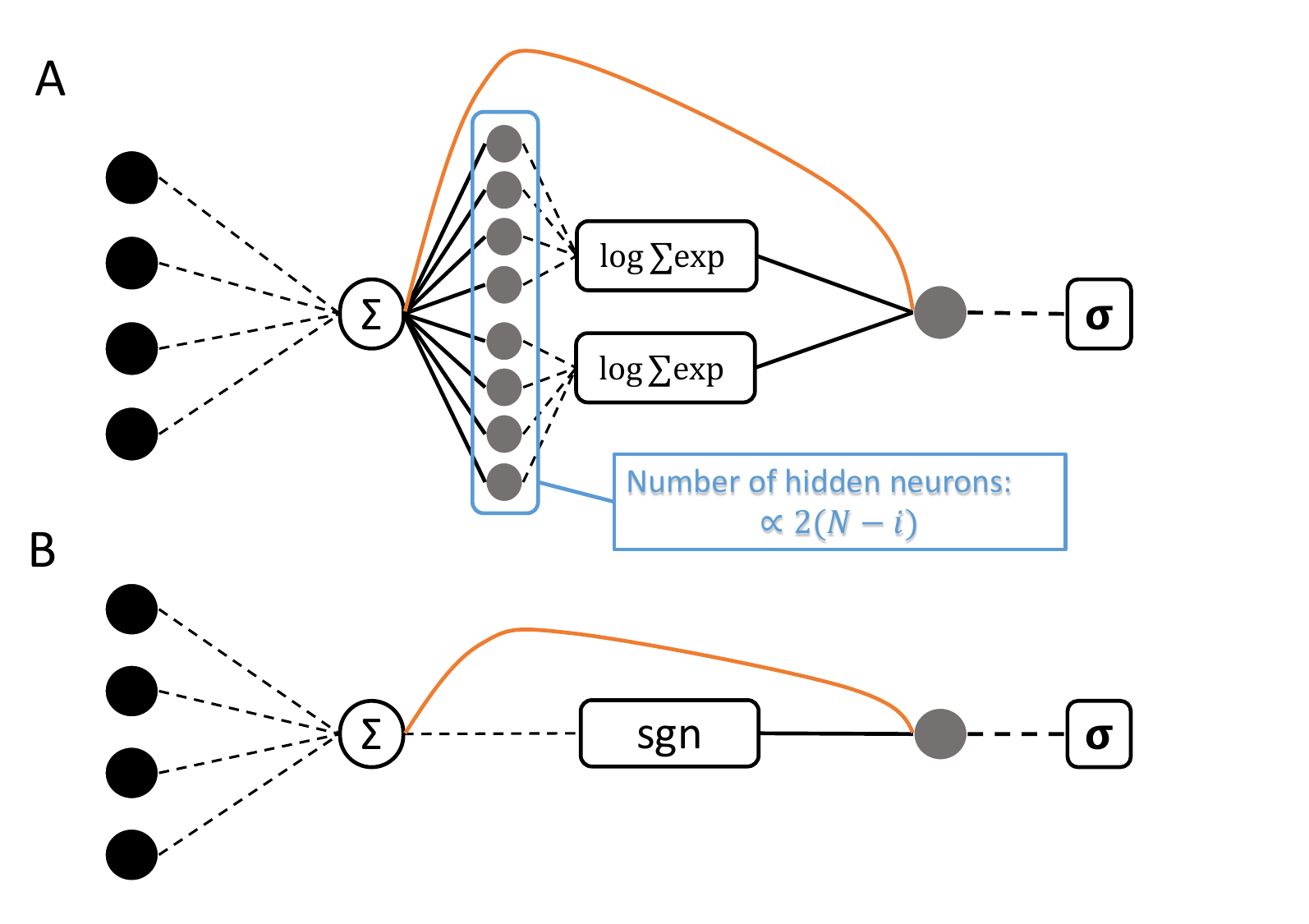}
    \caption{\textbf{CW$_N$ and CW$\infty$ architectures of a single conditional probability.}  Diagrams A and B represent the CW$_N$ and CW$\infty$ architectures, respectively. Both diagrams involve the operation of the sum of the input variables $\mathbf{x}_{<i}$. A skip connection, composed of a shared weight (represented by the orange line), is also present in both cases. In the CW$_N$ architecture, $2(N-1)$ linear operations are applied (with fixed weights and biases, as indicated in Eq.~(\ref{eq:x_i_first})), followed by two non-linear operations represented by $\log \sum \exp(x)$. On the other hand, in the CW$\infty$ architecture, apart from the skip connection, the input variables undergo a $sgn$ operation before being multiplied by a free weight parameter and passed through the final layer represented by the sigma function. The number of parameters in the CW$_N$ architecture scales as $2N^2$, while in the CW$\infty$ architecture, it scales as $N+1$.}
    \label{fig:CW_arch}
\end{figure}

The Curie-Weiss model (CW) is a uniform, fully-connected Ising model. The Hamiltonian, with $N$ spins, is $H\left(\mathbf{x}\right)=-h\sum_{i=1}^{N}x_{i}-\frac{J}{N}\sum_{i<j}x_{i}x_{j}$. The conditional probability of a spin $i$, Eq.~(\ref{eq:conditional_ghann}), of the CW model is:

\begin{multline}
P^{CW}\left(x_{i}=1|\mathbf{x}_{<i}\right) = 
\sigma\bigg( 
 2 \beta h + 2 \beta \frac{J}{N}\sum_{s=1}^{i-1}x_{s} + \\
 \log(\rho_i^+[\mathbf{x}_{<i}]) - \log(\rho_i^-[\mathbf{x}_{<i}])
\bigg),
\label{eq:conditional_cw}
\end{multline}
where:

\begin{equation}
\rho_i^{\pm}[\mathbf{x}_{<i}] \propto \sum_{\mathbf{x}_{>i}}e^{\beta \left(h\pm\frac{J}{N}+\frac{J}{N}\sum_{s<i}x_{s}\right)\sum_{l>i}x_{l}+\frac{\beta J}{2N}(\sum_{l,l'>i}x_{l} x_{l'})} 
\label{eq:rho_cw_0}
\end{equation}
Defining $h_i^{\pm}[\mathbf{x}_{<i}] =h\pm\frac{J}{N}+\frac{J}{N}\sum_{s=1}^{i-1}x_{s}$, at given $\mathbf{x}_{<i}$, Eq.~(\ref{eq:rho_cw_0}) is equivalent to the partition function of a CW model, with $N-i$ spins and external fields $h_i^{\pm}$. 
As shown in the Supporting Information (SI), the summations over $\mathbf{x_{>i}}$ can be easily done, finding the following expression:

 \begin{eqnarray}
 \rho_i^{\pm}[\mathbf{x}_{<i}] = \sum_{k=0}^{N-i} e^{b_{ik}^{\pm} + w_{ik} \sum_s x_s} 
\end{eqnarray}
where we defined:

\begin{align}
\label{eq:params}
\begin{split}
b_{ik}^{\pm} & = \log\binom{N-i}{k} + \frac{\beta J}{2N}\left(N-i-2k\right)^{2}+ \\
& \qquad \left(N-i-2k\right)\left(\beta h \pm \frac{\beta J}{N}\right)
\end{split} \\
\omega_{ik} & = \frac{\beta J}{N}\left(N-i-2k\right).
\label{eq:CW_params}
\end{align}
The final feed-forward architecture of the Curie-Weiss Autoregressive Neural Network (CW$_N$) architecture is:

\begin{multline*}
P^{CW_{N}}\left(x_{i}=+1|\mathbf{x}_{<i}\right)  =   \sigma \bigg[b+\omega \sum_{s=1}^{i-1}x_{s} \\
 + \log\big(\sum_{k=0}^{N-i}e^{b_{ik}^{+} + 
w_{ik}\sum_{s=1}^{i-1}x_{s}}\big)-\log\big(\sum_{k=0}^{N-i}e^{b_{ik}^{-} + w_{ik}\sum_{s=1}^{i-1}x_{s}}\big)\bigg],
\end{multline*}
where $b=2\beta h$, $\omega = \frac{2\beta J}{N}$ are the same, and so shared, among all the conditional probability functions, see Fig.~\ref{fig:CW_arch}. Their parameters have an analytic dependence on the parameters $J$ and $h$ of the Hamiltonian of the systems. 

The number of parameters of a single conditional probability of the CW$_N$ is $2+4(N-i)$, which decreases as $i$ increases. 
The total number of parameters of the entire conditional probability distribution scales as $2N^2$. 

If we consider the thermodynamical limit, $N \gg 1$, the ARNN architecture of the CW model simplifies (see sec.IB of the SI for details) to the following expression:

\begin{equation}
    \label{eq:curie_weiss_cond2}
    P^{CW_{\infty}}\left(x_{i}=1|\mathbf{x}_{<i}\right) =  \sigma \left(b+\omega \sum_{s=1}^{i-1}x_{s} + \omega_i^1 \text{sgn}(\sum_{s=1}^{i-1}x_{s})\right)
\end{equation}
where $b=2\beta h$, $\omega = \frac{2\beta J}{N}$ are the same as before, and shared, among all the conditional probability functions. The $\omega^1_i = -2\beta J |m_i|$ is different for each of them and can be computed analytically. 
The total number of parameters of the CW$_{\infty}$ scales as $N+2$.

\subsection{The Sherrington-Kirkpatrick (SK) model}
\label{sec:SK}
The SK Hamiltonian, considering zero external fields for simplicity, is given by:

\begin{equation}
H\left(\mathbf{x}\right)=-\sum_{i<j}J_{ij}x_{i}x_{j}
\end{equation}
where the set of couplings, $\underline{J}$, are i.i.d. random variable drawn from a Gaussian probability distribution $P(J)= \mathcal{N}(0, J^2/N)$. 

To find a feed-forward representation of the conditional probability of its Boltzmann distribution we have to compute the quantities in Eq.\ref{eq:rho_ghann}, that, defining $h_l^{\pm}[\mathbf{x}_{<i}] =\pm J_{il} + \sum_{s=1}^{i-1} J_{sl} x_s$, can be written as:

\begin{align*}
    \rho_i^{\pm} [\mathbf{x}_{<i}]  = \sum_{\mathbf{x}_{>i}}  \exp \bigg(
    \beta\sum_{l=i+1}^{N} h_l^{\pm}[\mathbf{x}_{<i}] x_l
    + \sum_{l'>l>i}^{N} J_{ll'} x_l x_{l'} \bigg)
\end{align*}
The above equation can be interpreted as an SK model over the variables $\mathbf{x}_{>i}$ with site-dependent external fields $h_l^{\pm}[\mathbf{x}_{<i}]$. 
I will use the replica trick \cite{10.1142/0271}, which is usually applied together with the average over the system's disorder. In our case, we deal with a single instance of disorder, with the set of couplings being fixed. In the following I will assume that $N-i \gg 1$, and the average over the disorder $\mathbb{E}$ is taken on the coupling parameters $J_{ll'}$ with $l,l'>i$. In practice, I will use the following approximation to compute the quantity:

\[
\log\rho_i^{\pm} \sim \mathbb{E}\left[  \log\rho_i^{\pm} \right] = \lim_{n\rightarrow 0} \frac{  \log(\mathbb{E}\left[(\rho_i^{\pm})^n \right])}{n}
\]
In the last equality, I use the replica trick. 
Implicitly, it is assumed that the quantities $\log\rho_i^{\pm}$ are self-averaged on the $\mathbf{x}_{>i}$ variables.
 The expression for the average over the disorder of the replicated function is:

\begin{multline}
\mathbb{E}_{\underline{J}_{ll'}}\left[(\rho_i^{\pm}[\mathbf{x}_{<i}])^n \right]  = 
\int \prod_{l<l'} dP_{J_{ll'}} \bigg\{ 
\sum_{\{\underline{x}^{a}\}_{i+1}^N} \exp\bigg[\\ \beta \bigg( 
\sum_{\substack{i<l \le N\\ 1<a<n}}h_l^{\pm}[\mathbf{x}_{<i}] x_l^{a} + 
\sum_{\substack{i < l< l' \le N\\ 1<a \le n}} J_{ll'} x_l^{a} x_{l'}^{a}
\bigg)  \bigg] 
\bigg\}
\end{multline}
where $dP_{J_{ll'}}=P(J_{ll'})dJ_{ll'}$, and the set of $\mathbf{x}^a$ are the replicated spin variables.
Computing the integrals over the disorder, we find: 

\begin{widetext}
\begin{align}
\mathbb{E}_{\underline{J}_{ll'}}\left[(\rho_i^{\pm}[\mathbf{x}_{<i}])^n \right]
& \propto  \int \prod_{a<b} dQ_{ab} e^{-\frac{N}{2}\beta^2Q_{a,b}^2}
\prod_{l} \left[
\sum_{\{\underline{x}^{a}_l\}} 
\exp\left\{\beta \left[
h_l^{\pm}[\mathbf{x}_{<i}] \sum_{a} x_l^{a} +\beta \sum_{a<b} Q_{ab}  x_l^{a} x_l^{b} \right]  \right\}
\right]
\end{align}
\end{widetext}
where in the last line I used the Hubbard-Stratonovich transformation to linearize the quadratic terms. See the SI or, for instance, \cite{nishimori2001statistical}, for details about the formal mathematical derivations of the previous and following expressions.
The Parisi solution of the SK model prescribes how to parametrize the matrix of the overlaps $\{Q_{ab}\}$ \cite{10.1142/0271}. The easiest way to parametrize the matrix of the overlaps is the replica symmetric solutions (RS), where the overlaps are equal and independent from the replica index: 

$$
Q_{ab}=\begin{cases}
			0, & \text{if $a=b$}\\
            q, & \text{otherwise}
		 \end{cases},
$$
A sequence of better approximations can then be obtained by breaking the replica symmetry step by step, from the 1-step replica symmetric breaking (1-RSB) to the k-step replica symmetric breaking (k-RSB) solution. The infinite k limit of the k-step replica symmetric breaking solution gives the exact solution of the SK model \cite{10.2307/20159953}.
The sequence of k-RSB approximations can be seen as nested non-linear operations \cite{Parisi_1980}, see the SI for details. 
\begin{figure}[!h]
    \centering 
    \includegraphics[width=0.48\textwidth]{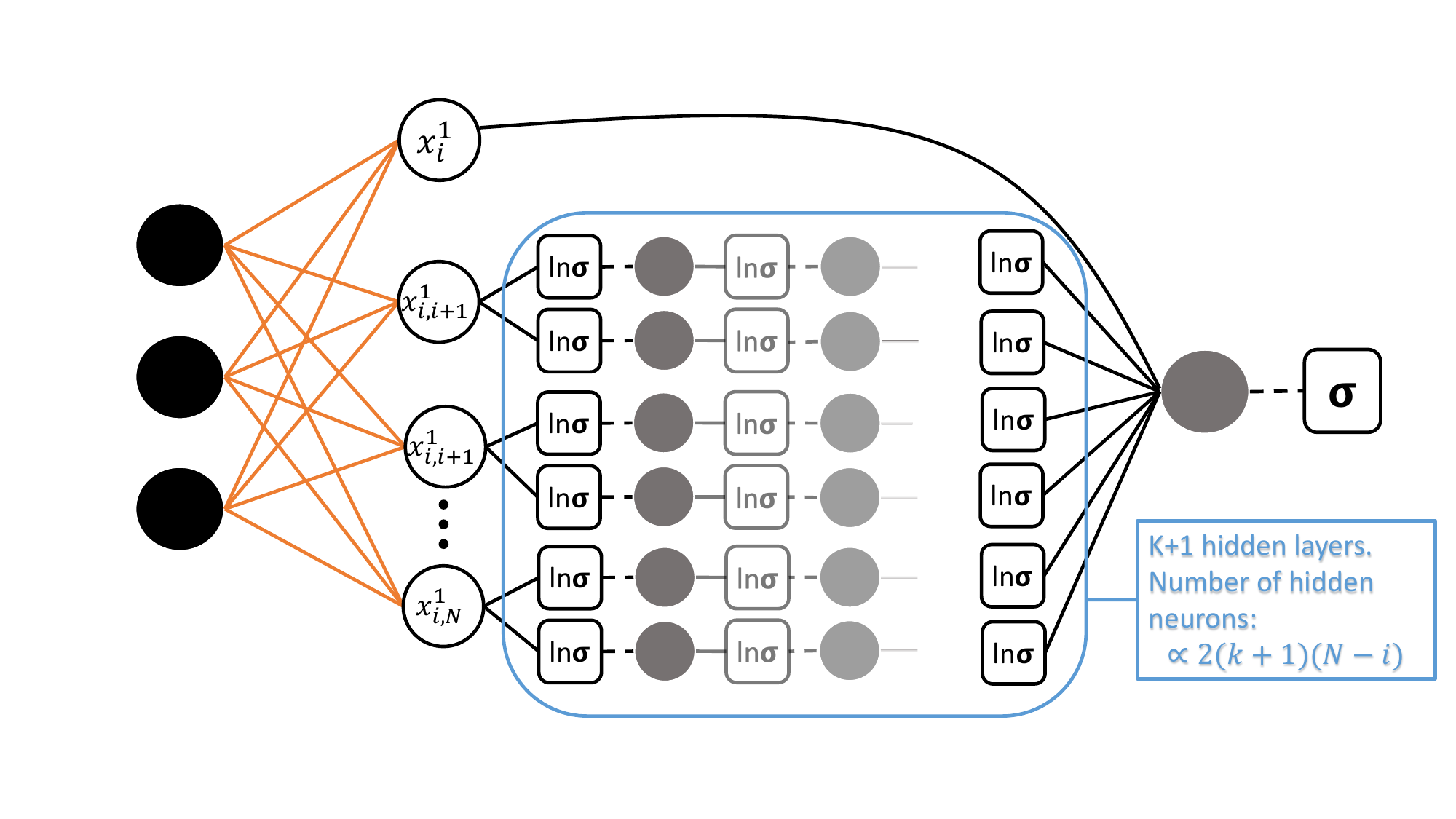}
    \caption{\textbf{SK$\mathbf{_{RS/kRSB}}$ architectures of the single variable conditional probability} The diagram depicts the SK$_{RS/kRSB}$ architectures that approximate a single conditional probability of the Boltzmann distribution in the SK model. The input variables are $\mathbf{x}_{<i}$, and the output is the conditional probability $Q^{\text{RS/k-RSB}}\left(x_{i}=1|\mathbf{x}_{<i}\right)$. The non-linear operations are represented by squares and the linear operations by solid lines. The parameters, in the orange lines, are equal to the Hamiltonian parameters and shared among the conditional probabilities, as indicated in Eq.~(\ref{eq:x_i_first}). The depth of the network is determined by the level of approximation used, with the $Q^{\text{RS}}$ architecture having only one hidden layer and the $Q^{\text{k-SRB}}$ architecture having a sequence of $k+1$ hidden layers. The total number of parameters scales as $2(k+1)N^2 + \mathcal{O}(N)$, where the $RS$ case corresponds to $k=0$.}
    \label{fig:SK_arch}
\end{figure}

Every k-step replica symmetric breaking solution leads to adding a Gaussian integral and two more free variational parameters to the representation of the $\rho^{\pm}$ functions. 
In the following, we will use a feed-forward representation that enlarges the space of parameters, using a more computationally friendly non-linear operator. 
Numerical evidence of the quality of the approximation used is shown in the SI.  
Overall, the parametrization of the overlaps matrix, which introduces free parameters in the derivation, allows the summing of all the configurations of the variables $\mathbf{x}_{i>}$ eliminating the exponential scaling with the system's size of the number of parameters.
The final ARNN architecture of the SK model is as follows (see the SI for details):

\begin{multline}
    Q^{\text{RS/k-RSB}}\left(x_{i}=1|\mathbf{x}_{<i}\right) = \sigma\bigg( 
        x_i^1(\mathbf{x}_{<i}) \\
        +\log(\rho_i^{+, \text{(RS/kRSB)}}) -
         \log(\rho_i^{-, \text{(RS/kRSB)}})
    \bigg). 
\end{multline}
For the RS and 1-RSB cases, we have:

\begin{align*}
    \log \rho^{\pm, RS} & = \sum_{l=i+1}^{N}  w_{il}^{0\pm} \log \sigma(b_{il}^{1{\pm}} +
w_{il}^{1{\pm}} x_{il}^1(\mathbf{x}_{<i})) \\
\begin{split}
    \log \rho^{\pm, 1RSB} & = 
    \sum_{l=i+1}^{N}  w_{il}^{0{\pm}} \log\sigma(b_{il}^{1{\pm}} + \\
    &  w_{il}^{1{\pm}} \log\sigma(b_{il}^{2{\pm}} +  w_{il}^{2{\pm}}  x_{il}^1(\mathbf{x}_{<i}))). 
    \end{split}
\end{align*}
The set of $x_{il}^1(\mathbf{x}_{<i})$ is the output of the first layer of the ARNN, see eqs.\ref{eq:x_i_first}-\ref{eq:x_il_first}, and $(w_{il}^{0{\pm}}, b_{il}^{1{\pm}}, w_{il}^{1{\pm}}, b_{il}^{2{\pm}}, w_{il}^{2{\pm}})$ are free variational parameters of the ARNN (see Fig.~\ref{fig:SK_arch}). The number of parameters of a single conditional probability distribution scales as $2(k+1)(N-i)$ where $k$ is the level of the k-RSB solution used, assuming $k=0$ as the RS solution.

\section{Results}

In this section, I compare several ARNN architectures in learning to generate samples from the Boltzmann distribution of the CW and SK models. 
Additionally, the ability to recover the Hamiltonian coupling parameters from Monte-Carlo-generated instances is presented. 
The CW$_N$, CW$_{\infty}$ and SK$_{RS/kRSB}$ architectures, presented in previous sections, are compared with: 

\begin{itemize}
    \item The one parameter (1P) architecture, where a single weight parameter is multiplied by the sums of the input variables, and then the sigma function is applied. This architecture was already used for the CW system in \cite{https://doi.org/10.48550/arxiv.2210.11145}. The total number of parameters scales as $N$.
    \item The single layer (1L) architecture, where a fully connected single linear layer parametrizes the whole probability distribution, where a mask is applied to a subset of the weights in order to preserve the autoregressive properties. The width of the layer is $N$, and the total number of parameters scale as $N^2$ \cite{pmlr-v37-germain15}.
    \item The MADE architecture\cite{pmlr-v37-germain15}, where the whole probability distribution is represented with a deep sequence of fully connected layers, with non-linear activation functions and masks in between them, to assure the autoregressive properties. Respect the 1L the deep architecture of MADE enhances the expressive power. The MADE$_{dc}$ used has $d$ hidden layers, each of them with $c$ channels of width $N$. For instance, the 1L architecture is equivalent to the MADE$_{11}$ and MADE$_{23}$ has two hidden fully-connected layers, each of them composed of three channels of width $N$. 
\end{itemize}

The parameters of the ARNN are trained to minimize the Kullback-Leibler divergence or, equivalently, the variational free energy (see Eq.~(\ref{eq:kl})). Given an ARNN, $Q^{\theta}$, that depends on a set of parameters $\theta$ and the Hamiltonian of the system $H$, the variational free energy can be estimated as:
\begin{align*}
F[Q^{\theta}] & = \sum_{\left\{ \mathbf{x} \right\}}Q^{\theta}\left[\frac{1}{\beta}\log Q^{\theta} + H[\mathbf{x}] \right]\\
& \approx \sum_{\mathbf{x} \sim Q^{\theta}} \left[ \frac{1}{\beta}\log Q^{\theta} + H[\mathbf{x}]\right].
\end{align*}
The samples are drawn from the trial ARNN, $Q^{\theta}$, using ancestral sampling. At each step of the training, the derivative of the variational free energy with respect to the parameters $\theta$ is estimated and used to update the parameters of the ARNN. Then a new batch of samples is extracted from the ARNN and used again to compute the derivative of the variational free energy and update the parameters\cite{Wu2019}. This process was repeated until a stop criterion is met or a maximum number of steps is reached. For each model and temperature, a maximum $1000$ epochs are allowed, with a batch size of $2000$ samples, and a learning rate of $0.001$. The ADAM algorithm\cite{kingma2014adam} was applied for the optimization of the ARNN parameters. An annealing procedure was used to improve performance and avoid mode-collapse problems\cite{Wu2019}, where the inverse temperature $\beta$ was increased from $0.1$ to $2.0$ in steps of $0.05$. The code was written using the PyTorch framework\cite{NEURIPS2019_bdbca288}, and it is open-source, released in \cite{mygithub}.
The CW$_N$ has all its parameters fixed and precomputed analytically, see Eq.~(\ref{eq:params}). The CW$_{\infty}$ has one free parameter for each of its conditional probability distributions to be trained, and one shared parameter, see Eq.~(\ref{eq:curie_weiss_cond2}). The parameters of the first layer of the SK$_{RS/kRSB}$ architecture are shared and fixed by the values of the couplings and fields of the Hamiltonian. The parameters of the hidden layers are free and trained. The parameters of the MADE$_{dc}$, 1L and 1P architectures are free and trained.
As explained in Sec. \ref{sec:ARNN_boltzmann}, the variational free energy $F[Q^{\theta}]$ is always an upper bound of the free energy of the system. Its value will be used, in the following, as a benchmark for the performance of the ARNN architecture in approximating the Boltzmann distribution. After the training procedure, the variational free energy was estimated using 20,000 configurations sampled from each of the considered ARNN architectures. 
The training procedure was the same for all the experiments unless conversely specified.
\subsection{The CW model}
\begin{figure}[h]
    \centering
    \includegraphics[width=0.45\textwidth]{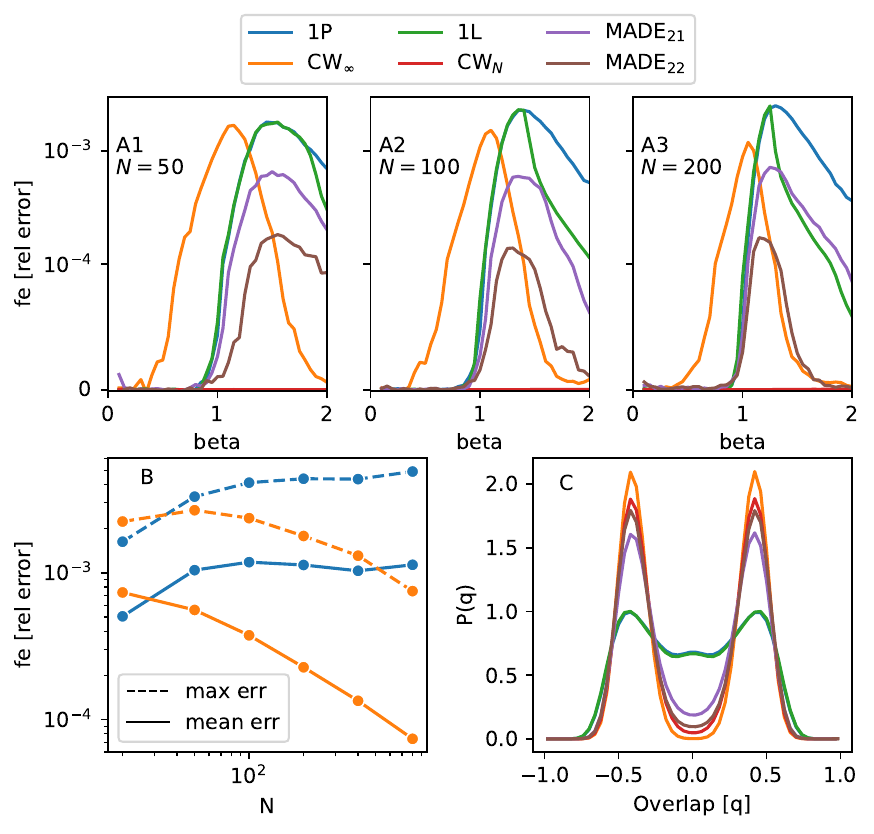}
    \caption{\textbf{Results for CW model.} The CW model considered has $J=1$ and $h=0$ (see the text for details). The system undergoes a second-order phase transition at $\beta=1$ where a spontaneous magnetization appears\cite{kadanoff2000statistical}. [\textbf{A1, A2, A3}] Relative error in the estimation of the free energy for different system sizes with respect to the analytic solution. The CW$_N$ architecture has its parameters fixed and precomputed analytically, and the error is too small to be seen at this scale. The y-axis is plotted on a logarithmic scale down to $10^{-4}$  and then linearly to zero.[\textbf{B}] The dependence on $N$ of the mean and maximum relative error of the two smaller architectures, $1P$ and CW$_{\infty}$, both of which scale linearly with the size of the system. [\textbf{C}] Distribution of the overlaps of the samples generated by the ARNNs for the CW system with $N=200$ variables and $\beta=1.3$}
    \label{fig:curie_weiss}
\end{figure}

The results on the CW model, with $J=1$ and $h=0$, are shown in Fig.~\ref{fig:curie_weiss}. 
The plots A1, A2, and A3, in the first row, show the relative error of the free energy density ($fe[P] = F[P]/N$), with respect to the exact one, computed analytically \cite{kadanoff2000statistical}, see the SI for details, for different system sizes $N$.  
The variational free energy density estimated from samples generated with the CW$_N$ architecture does not have an appreciable difference with the analytic solution, and for the CW$_{\infty}$, it improves as the system size increases. Fig.~\ref{fig:curie_weiss}.B plots the error, in the estimation of the free energy density for the architectures with fewer parameters, 1P and CW$_{\infty}$ (both scaling linearly with the system's size); It shows clearly that a deep architecture with skip connections, in this case with only one more parameter, in the skip connection, improves the accuracy by orders of magnitude. The need for deep architectures, already on a simple model as the CW, is indicated by the poor performance of the 1L architecture, despite its scaling of parameters as $N^2$, it achieves similar results to the 1P. The MADE architecture obtains good results but was not comparable to CW$_N$, even though having a similar number of parameters. The plot in Fig.~\ref{fig:curie_weiss}.c shows the distribution of the overlaps, $q_{\mathbf{a}, \mathbf{b}}=\frac{1}{N}\sum_{i} a_i b_i$ where $a_i, b_i$ are two system configurations, between the samples generated by the ARNNs. The distribution is computed at $\beta=1.3$ for $N=200$. It can be seen as the poor performance of the 1-layer networks (1P, 1L) is due to the difficulty of correctly representing the configurations with magnetization different from zero in the proximity of the phase transition. This could be due to mode collapse problems \cite{https://doi.org/10.48550/arxiv.2210.11145}, which do not affect the deeper ARNN architectures tested.
\subsection{The SK model}
\begin{figure}[h]
    \centering 
    \includegraphics[width=0.48\textwidth]{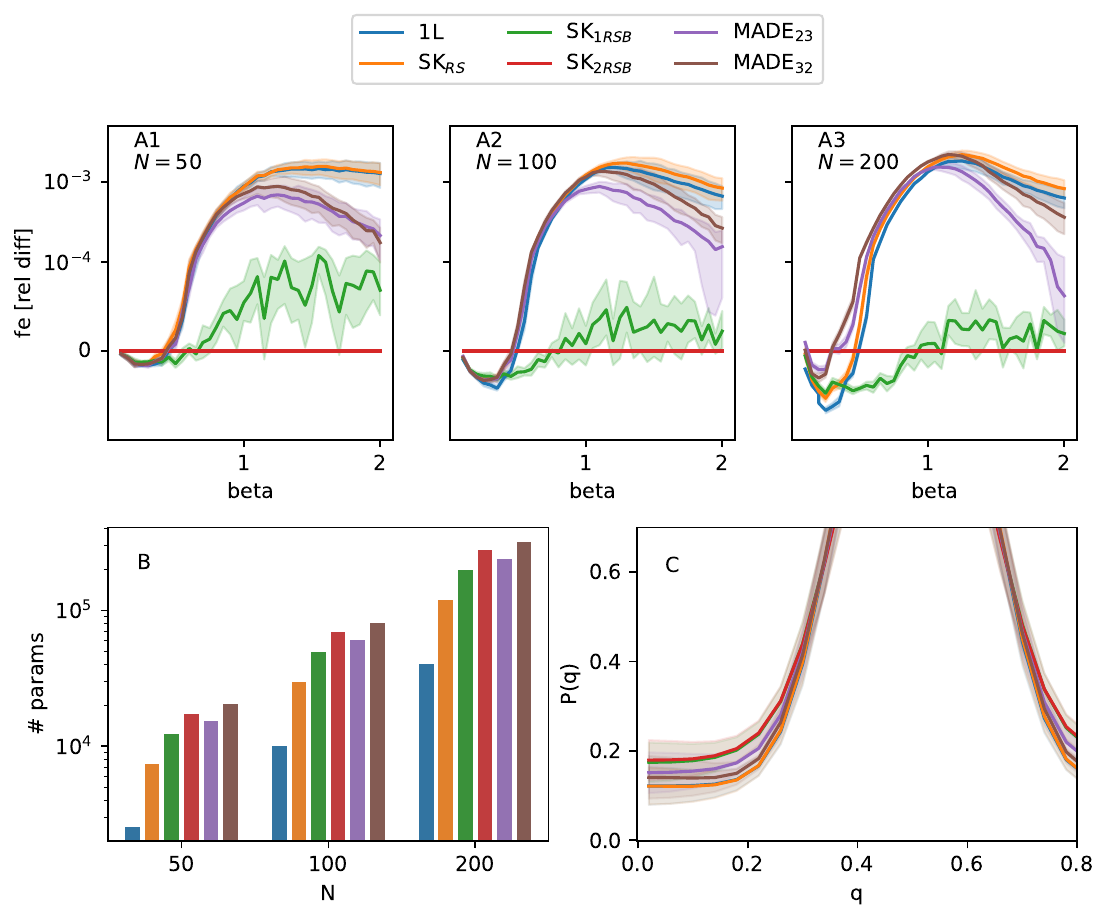}
    \caption{\textbf{Results for SK model.} The SK model considered has $J=1$ and $h=0$ (see the text for details). The system undergoes a phase transition at $\beta=1$\cite{10.1142/0271}. [\textbf{A1, A2, A3}] Relative difference in the estimation of the free energy for increasing system sizes with respect to the free energy computed by SK$_{2RSB}$ architecture. The results are averaged over 10 instances of the disorder. The y-axis is plotted on a logarithmic scale down to $10^{-4}$  and then linearly to $-10^4$. [\textbf{B}] Scaling with $N$ of the number of parameters of the ARNN architectures. [\textbf{C}] Distribution of the overlaps of the samples generated by the ARNNs architectures for the SK model with $N=200$ variables and $\beta=1.5$, averaged over 10 different instances. The translucent error bands surrounding the plotted lines represent the $95\%$ confidence intervals.}
    \label{fig:SK}
\end{figure}

In Fig.~\ref{fig:SK}, the result of the SK model, with $J=1$ and $h=0$ are shown; as before in the first row there is the relative error in the estimation of the free energy density at different system sizes. 
In this case, the exact solution, for a single instance of the disorder and a finite $N$ is not known. The free energy estimation of the SK$_{2RSB}$ was taken as the reference to compute the relative difference. The free energy estimations of SK$_{kRSB}$ with $k=1,2$ are very close to each other.
The performance of the SK$_{RS}$ net is the same as the 1L architecture even with a much higher number of parameters. The MADE architecture tested, even with a similar number of parameters of the SK$_{kRSB}$ nets, see Fig.~\ref{fig:SK}.C, estimate a larger free energy, with differences increasing with $N$.
To better assess the difference in the approximation of the Boltzmann distribution of the architecture tested, I consider to check the distributions of the overlaps $q$ among the generated samples. The SK model, with $J=1$ and $h=0$, undergoes a phase transition at $\beta=1$, where a glassy phase is formed, and an exponential number of metastable states appears \cite{10.1142/0271}. This fact is reflected in the distribution of overlaps that have values different from zero in a wide region of values of $q$ \cite{PhysRevLett.51.1206}.
Observing the distribution of the overlaps in the glassy phase, $\beta=1.3$, between the samples generated by the ARNNs, Fig.~\ref{fig:SK}.C, we can check as the distribution generated by the SK$_{kRSB}$ is higher in the region between the peak and zero overlaps, suggesting that these architectures better capture the complex landscape of the SK Boltzmann probability distribution \cite{PhysRevLett.51.1206}.

\begin{figure}[h]
    \centering 
    \includegraphics[width=0.45\textwidth]{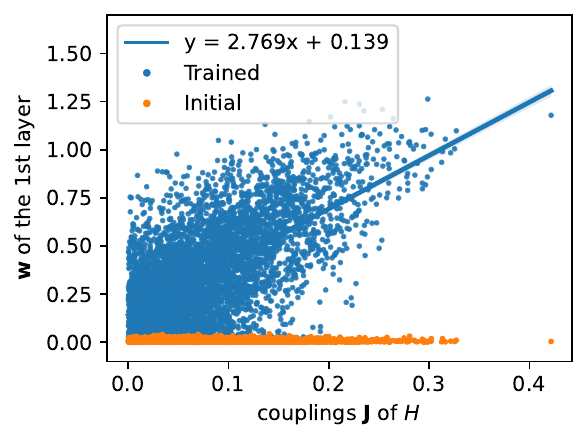}
    \caption{\textbf{Scatter plot of the weights vs the couplings.} Scatter plot of the absolute values of weights of the first layer of a SK$_{1RSB}$ vs the absolute values of the coupling parameters of the SK model. The weights are trained over 10,000 samples generated by the Metropolis Monte Carlo algorithm on a single instance of the SK model with $N=100$ variables at $\beta=2$. They are initialized at small random values. The blue line is the fit of the blue points, clearly showing a strong correlation between the weights and the coupling parameters of the Hamiltonian.}
    \label{fig:SK_MC}
\end{figure}

The final test for the derived SK$_{kRSB}$ architectures involves evaluating the ability to recover the Hamiltonian couplings of the system using only samples extracted from the Boltzmann distribution of a single instance of the SK model in the glassy phase at $\beta=2$. The Metropolis Monte-Carlo algorithm was used to sample, every 200 Monte-Carlo sweeps, 10,000 system configurations. The SK$_{1RSB}$ was trained to minimize the log-likelihood computed on these samples (see the SI for details). According to the derivation of the SK$_{kRSB}$ architecture, the weights of the first layer of the neural network should correspond to the coupling parameters of the Hamiltonian. Due to the gauge invariance of the Hamiltonian with respect to the change of sign of all the couplings $J$s, I will consider their absolute values in the comparison. The weights parameters of the first layers of the SK$_{1RSB}$ were initialized at small random values. As shown in Fig.~\ref{fig:SK_MC}, there is a strong correlation between the weights of the first layer and the couplings of the Hamiltonian, even though the neural network was trained in an over-parameterized setting; it has 60,000 parameters, significantly more than the number of samples.

\section{Conclusions}
In this study, the exact architecture  Autoregressive Neural Network (ARNN) architecture (H$_2$ARNN) of the Boltzmann distribution of the pairwise interacting system Hamiltonian was derived. The H$_2$ARNN is a deep neural network, with the weights and biases of the first layer corresponding to the couplings and external fields of the Hamiltonian, see eqs.\ref{eq:x_i_first}-\ref{eq:x_il_first}. The  H$_2$ARNN architecture has skip-connections and a recurrent structure with a clear physical interpretation. Although the H$_2$ARNN is not directly usable due to the exponential increase in the number of hidden layer parameters with the size of the system, its explicit formulation allows using statistical physics techniques to derive tractable architectures for specific problems. For example, ARNN architectures, scaling polynomially with the system's size, are derived for the CW and SK models. In the case of the SK model, the derivation is based on the sequence of k-step replica symmetric breaking solutions, which were mapped to a sequence of deeper ARNNs architectures.

The results, checking the ability of the ARNN architecture to learn the Boltzmann distribution of the CW and SK models, indicate that the derived architectures outperform commonly used ARNNs. Furthermore, the close connection between the physics of the problem and the neural network architecture is shown in the results of Fig.~\ref{fig:SK_MC}. In this case, the  SK$_{1RSB}$ architecture was trained on samples generated with the Monte-Carlo technique from the Boltzmann distribution of an SK model; the weights of the first layer of the SK$_{1RSB}$ were found to have a strong correlation with the coupling parameters of the Hamiltonian.

Even though the derivation of a simple and compact ARNN architecture is not always feasible for all types of pairwise interactions and exactly solvable physics systems are rare, the explicit form of the H$_2$ARNN and its clear physical interpretation provides a means to derive approximate architectures for specific Boltzmann distributions. 

In this work, while the ARNN architecture of an SK model was derived, its learnability was not thoroughly examined. The problem of finding the configurations of minimum energy for the SK model is known to belong to the NP-hard class, and the effectiveness of the ARNN approach in solving this problem is still uncertain and a matter of ongoing research \cite{https://doi.org/10.48550/arxiv.2210.11145, 10.1038/s42256-021-00401-3, condmat7020038}. Further systematic studies are needed to fully understand the learnability of the ARNN architecture presented in this work at very low temperatures and also on different systems.

There are several promising directions for future research to expand upon presented ARNN architectures. For instance, deriving the architecture for statistical models with more than binary variables. In statistical physics, the models with variables that have more than two states are called Potts models. The language models, where each variable represents a word, and could take values among a huge number of states, usually more than tens of thousand possible words (or states), belong to this set of systems. The generalization of the present work to Potts models could allow us to connect the physics of the problem to recent language generative models like the transformer architecture \cite{rende2023optimal}. Another direction could be to consider systems with interactions beyond pairwise, to describe more complex probability distributions. Additionally, it would be interesting to examine sparse interacting system graphs, such as systems that interact on grids or random sparse graphs. The first case is fundamental for a large class of physics systems and image generation tasks, while the latter type, such as Erdos-Renyi interaction graphs, is common in optimization \cite{doi:10.1126/science.1073287} and inference problems \cite{Biazzo2022}. 

\section*{Acknowledgments}
I.B. thanks Giuseppe Carleo, Giovanni Catania, Vittorio Erba, Guido Uguzzoni, Martin Weigt, Francesco Zamponi, Lenka Zdeborová and all the members of the SPOC-lab of the EPFL for useful discussions. I.B. thanks also Christian Keup for reading the manuscript and for his valuable comments.

\section*{Author contributions statement}
I.B. conceived the idea, wrote the code, run the simulations, analyzed the data, and wrote the manuscript.

\section*{Competing interests}
The author declares no competing interests.

\section*{Data availability}
All data and code needed to reproduce the results are released on GitHub:  \href{https://github.com/ocadni/h2arnn}{h2arnn}.

\bibliography{main}

\end{document}


\title{Supplementary information of: \emph{``The Autoregressive Neural Network Architecture of the Boltzmann Distribution of Pairwise Interacting Spin Systems"}}

\author{Indaco Biazzo}
\maketitle

\section{ANRR architecture of the Curie-Weiss model}

The Curie-Weiss model (CW) is a uniform, fully-connected Ising model. The Hamiltonian of the Curie-Weiss model (CW) with $N$ spins is $H\left(\mathbf{x}\right)=-h\sum_{i=1}^{N}x_{i}-\frac{J}{N}\sum_{i<j}x_{i}x_{j}$. Defining $M_{<i}[\mathbf{x}_{<i}]=\sum_{s=1}^{i-1}x_{s}$ and $M_{>i}=\sum_{l=i+1}^{N}x_{l}$, the conditional probability of a spin $i$, see Eq.~(11) in the main text, of the CW model, is:

\begin{equation}
P^{CW}\left(x_{i}=1|\mathbf{x}_{<i}\right) = 
\sigma\bigg( 
 2 \beta h + 2 \beta \frac{J}{N}M_{i<}[\mathbf{x}_{<i}] + 
 \log(\rho_i^+[\mathbf{x}_{<i}]) - \log(\rho_i^-[\mathbf{x}_{<i}])
\bigg),
\label{eq:conditional_cw}
\end{equation}

where:

\begin{equation}
\rho_i^{\pm}[\mathbf{x}_{<i}] \propto \sum_{\mathbf{x}_{>i}}e^{\beta \left(h\pm\frac{J}{N}+\frac{J}{N}M_{<i}[\mathbf{x}_{<i}]\right)M_{>i}+\frac{\beta J}{2N}M_{>i}^{2}} 
\label{eq:rho_cw_0}
\end{equation}

Defining $h_i^{\pm}[\mathbf{x}_{<i}] =h\pm\frac{J}{N}+\frac{J}{N}M_{<i}[\mathbf{x}_{<i}]$, at given $\mathbf{x}_{<i}$, the Eq.~(\ref{eq:rho_cw_0}) is equivalent to the partition function of the CW model, with $N-i$ spins and external fields $h_i^{\pm}$, as written in the main text. 
The sums over the configurations of the spins $l$ can be carried on easily: 

\begin{multline}
 \rho_i^{\pm}[\mathbf{x}_{<i}] = \sum_{x_{i+1}\dots x_{N}} e^{\beta h_i^{\pm}[\mathbf{x}_{<i}]M_{>i} +\frac{\beta J}{2N}M_{>i}^{2}} = \sqrt{\frac{N}{2\pi \beta J}}\sum_{x_{i+1}\dots x_{N}}e^{\beta h_i^{\pm}[\mathbf{x}_{<i}] M_{>i}}\int e^{-\frac{N}{2J \beta}t^{2}+t M_{>i}} dt\\
  = \sqrt{\frac{N}{2\pi \beta J}}\int dt e^{-\frac{N}{2J \beta}t^{2}} \sum_{x_{i+1}\dots x_{N}}e^{(\beta h_i^{\pm}[\mathbf{x}_{<i}] + t) M_{>i}}  
 =  \sqrt{\frac{N}{2\pi \beta J}}\int dt e^{-\frac{N}{2J \beta}t^{2}} \left(e^{\beta h_i^{\pm}[\mathbf{x}_{<i}] + t} + e^{ (-\beta h_i^{\pm}[\mathbf{x}_{<i}] - t)} \right)^{N-i}  \\ 
 \label{eq:rho_last_exact}
 \end{multline} 

 where we used the Hubbard–Stratonovich (HS) transformation to obtain the second equality. The last equality is obtained just by summing over the variables $x_l$.

 First, in the following, we derive the exact feed-forward representation of Eq.~(\ref{eq:conditional_cw}) at finite $N$ number of variables, then in the limit of $N\rightarrow \infty$.

 \subsection{Exact expression of the conditional probability of the CW model}
 The integral in the equation \ref{eq:rho_last_exact} can be computed the following way:

 \begin{eqnarray*}
 \rho_i^{\pm}[\mathbf{x}_{<i}] &=& \sqrt{\frac{N}{2\pi \beta J}}\int dt e^{-\frac{N}{2J \beta}t^{2}} 
 \sum_{k=0}^{N-i} \binom{N-i}{k} e^{(N-i-2k)*(\beta h_i^{\pm}[\mathbf{x}_{<i}] + t)}\\
 &=& \sum_{k=0}^{N-i} \binom{N-i}{k} \sqrt{\frac{N}{2\pi \beta J}}\int dt e^{-\frac{N}{2J \beta}t^{2}} 
  e^{(N-i-2k)*(\beta h_i^{\pm}[\mathbf{x}_{<i}] + t)}\\
&=& \sum_{k=0}^{N-i} \binom{N-i}{k}e^{\frac{\beta J}{2N}\left(N-i-2k\right)^{2}+\left(N-i-2k\right)\left(\beta h \pm \frac{\beta J}{N}\right)} e^{\frac{\beta J}{N}\left(N-i-2k\right) \sum_s x_s} \\
&=& \sum_{k=0}^{N-i} e^{b_{i,k}^{\pm} + w_{i,k} \sum_s x_s} 
\end{eqnarray*}

where:

\begin{eqnarray}
\label{eq:params}
b_{ik}^{\pm} & = & \log\binom{N-i}{k} + \frac{\beta J}{2N}\left(N-i-2k\right)^{2}+\left(N-i-2k\right)\left(\beta h \pm \frac{\beta J}{N}\right)\\
\omega_{ik} & = & \frac{\beta J}{N}\left(N-i-2k\right),
\end{eqnarray}

proving Eq.~(10) shown in the main text.

\subsection{Thermodynamical limit of the conditional probability of the CW model}
In the thermodynamical limit, the Curie-Weiss model admits an analytical solution. The order parameter of the system is the magnetization, $m_{\beta}=\frac{1}{N Z}\sum_{\mathbf{x}}\sum_i x_i e^{-\beta H}$ with $Z = \sum_{\mathbf{x}} e^{-\beta H}$. At high temperatures, with zero external fields $h=0$, the magnetization, $m_{\beta}$, is zero up to a critical temperature $\beta_c=\frac{1}{J}$, where a phase transition occurs, and a non-zero magnetization is observed \cite{kadanoff2000statistical}. Considering the following variables: $\eta_i = \frac{N-i}{N}$, $m_i = -\frac{N-i-2k}{N}$, and for simplicity, the $h=0$ case, we can rewrite the expression, Eq.~(\ref{eq:rho_last_exact}), as:

\begin{align}
    \rho_i^{\pm}[\mathbf{x}_{<i}] &= \sqrt{\frac{N}{2\pi \beta J}}\int dt e^{-\frac{N}{2J \beta}t^{2}} 
    \sum_{k=0}^{N-i} \binom{N-i}{k} e^{(N-i-2k)*(\beta h_i^{\pm}[\mathbf{x}_{<i}] + t)}\\ \label{eq:rho_last_exact2} &= \sum_{k=0}^{N-i} \binom{N-i}{k}e^{\frac{\beta J}{2N}\left(N-i-2k\right)^{2}+\left(N-i-2k\right)\left(\pm\frac{\beta J}{N}\right)} e^{\frac{\beta J}{N}\left(N-i-2k\right) \sum_s x_s}  \\
    &= \sum_{m_i=-\eta_i}^{\eta_i} \binom{N \eta_i}{\frac{N(m_i+\eta_i)}{2}} e^{\frac{N \beta J}{2}m_i^{2} \mp \beta J m_i } e^{- N \beta J \frac{\sum_s x_s}{N}}
\end{align}    

In the limit $N \gg 1$, we obtain:

 \begin{align}
 \rho_i^{\pm} & = 
  \int_{-\eta_i}^{\eta_i} \binom{N\eta_i}{\frac{N(m_i+\eta_i)}{2}} e^{\frac{N \beta J}{2}m_i^{2} \mp \beta J m_i } e^{N \beta J \frac{\sum_s x_s}{N}} dm_i \\
\begin{split} 
  \approx & \int_{-\eta_i}^{\eta_i} \exp\bigg\{-N\eta_i \big( -\frac{1+\frac{m_i}{\eta_i}}{2} \log\frac{1+\frac{m_i}{\eta_i}}{2} \\
   & - \frac{1-\frac{m_i}{\eta_i}}{2} \log\frac{1-\frac{m_i}{\eta_i}}{2} 
      - \frac{\beta m_i^2}{2 \eta_i} + \beta m_i \frac{\sum_s x_s}{N}\big) \bigg\} e^{\mp \beta J m_i} dm_i
\end{split} \\
\begin{split} 
  \approx & \exp\bigg\{-N\eta_i \big( -\frac{1+\frac{\tilde{m}_{i}}{\eta_i}}{2} \log\frac{1+\frac{\tilde{m}_{i}}{\eta_i}}{2} \\
   & - \frac{1-\frac{\tilde{m}_{i}}{\eta_i}}{2} \log\frac{1-\frac{\tilde{m}_{i}}{\eta_i}}{2} 
      - \frac{\beta \tilde{m}_{i}^2}{2 \eta_i} + \beta \tilde{m}_{i} \frac{\sum_s x_s}{N}\big) \bigg\} e^{\mp \beta J \tilde{m}_{i}} \\
    \doteq & e^{\mp \beta J \tilde{m}_{i}} 
\end{split}
\end{align}

where in the second line I used the Stirling approximation for the binomial factor, in the third line I used the saddle point method, assuming $N\gg 1$, where $\tilde{m}_{i}$ is the extreme of the function inside the curly brackets. In the last line, I simplify all the elements that are equal between $\rho_i^{+} $ and $\rho_i^{-}$ that cancel each other in the conditional probability function Eq.~(\ref{eq:conditional_cw}). Deriving by $m_i$ the function inside the curly brackets, we obtain that the extreme $\tilde{m}_{i}$ should satisfy the following equation:

\begin{equation}
\frac{\tilde{m}_{i}}{\eta_i} = \tanh \left( \beta(\frac{\tilde{m}_{i}}{\eta_i} - \frac{\sum_s x_s}{N}) \right)
\label{eq:extrem_i}
\end{equation}
In the $N$ large limit, and for a typical sample, I assume that: $ \frac{\sum_s x_s}{N} = \frac{i}{N} \frac{\sum_s x_s}{i}  \approx \frac{i}{N} |\tilde{m}_{\beta}| \text{sgn}(\sum_s x_s)$, where the $m_{\beta}$ is the magnetization of the Curie-Weiss system at inverse temperature $\beta$ and $\text{sgn(x)} = \frac{|x|}{x}$ is the sign function.
We can distinguish two cases when the magnetization of the system is zero or not. 
In the first case, when $\beta\leq 1/J$, the solution of Eq.~(\ref{eq:extrem_i}) is zero as well, and $\log(\rho_i^{+}) - \log(\rho_i^{-})=0$ because the only term that makes the two quantities different, $\mp \beta J m_i$, vanish.\\ 
When instead the system acquires a magnetization $m_{\beta}$ different from zero, Eq.~(\ref{eq:extrem_i}) admit one maximum that depends on the two possible symmetric values of $\frac{i}{N} \frac{\sum_s x_s}{i}\approx \frac{i}{N} |\tilde{m}_{\beta}| \text{sign}(\sum_q x_q)$. 
The solution of Eq.~(\ref{eq:extrem_i}), $\pm \tilde{m}_{\text{extrem}}$ depends again on $\text{sign}(\sum_s x_s)$, and we can write the maximum solution as $\tilde{m}_{i}=|\tilde{m}_i| \text{sgn}(\sum_s x_s)$. 
Easily we obtain that $\log(\rho_i^{+}) - \log(\rho_i^{-}) = -2\beta J|\tilde{m}_i| \text{sgn}(\sum_s x_s)$, demonstrating the formula in the main text.

\section{Analytic expression of the Curie-Weiss free energy at finite N}
The free energy of a system is defined as:

\begin{equation}
F = -\frac{1}{\beta} \log Z
\end{equation}
For the CW system, with similar computation that leads to  Eq.~(\ref{eq:rho_last_exact2}), we find:

\begin{equation}
F = -\frac{1}{\beta}\log \sum_{k=0}^{N} \binom{N}{k}e^{\frac{\beta J}{2}\left(N-2k\right)^{2}+\left(N-2k\right)h}  + \frac{J}{2}
\end{equation}

\section{ARNN architectures of The Sherrington-Kirkpatrick model}
In order to derive the architecture for the SK model based on the replica method, I'll start from Eq.~(17) presented in the main text:

\begin{equation}
\mathbb{E}_{\underline{J}}\left[(\rho_i^{\pm}[\mathbf{x}_{<i}])^n \right]  = \\
\int \prod_{l<l'} dP_{J_{ll'}} \bigg\{ 
\sum_{\{\underline{x}^{a}\}_{i+1}^N} \exp\bigg[\beta \bigg(
\sum_{l,a}\bigg( \pm J_{il} + \sum_{s} J_{sl} x_s \bigg) x_l^{a} + 
\sum_{l,l', a} J_{ll'} x_l^{a} x_{l'}^{a}
\bigg)  \bigg] 
\bigg\}
\end{equation}
where, the sums over $(l,l')$, $s$ and $a$ run respectively over $(i+1,N)$, $(1,i-1)$ and $(1,n)$, and $dP_{J_{ll'}}=P(J_{ll'})dJ_{ll'}$. Here we assumed that $N-i \gg 1$, and the average over the disorder $\mathbb{E}_{\underline{J}}$ is on the parameters $J_{l,l'}$ with $l,l'>i$. Defining $h_l^{\pm}[\mathbf{x}_{<i}] =\pm J_{il} + \sum_{s=1}^{i-1} J_{sl} x_s$ as an external field, we can observe that the above quantity is the partition function of the SK model on $\mathbf{x}_{i>}$ variables, with external fields $h_l^{\pm}[\mathbf{x}_{<i}]$, at fixed $\mathbf{x}_{<i}$, and $J_{ll'}$ as coupling constants. \\  
Computing the integrals over the disorder variables $\{J_{ll'}\}$ yields \cite{nishimori2001statistical}:

\begin{widetext}
\begin{eqnarray}
\mathbb{E}_{\underline{J}}\left[(\rho_i^{\pm}[\mathbf{x}_{<i}])^n \right] & = & 
\sum_{\{\underline{x}^{a}\}_{i+1}^N} 
\exp\left\{\beta \left[
\sum_{l} h_l^{\pm}[\mathbf{x}_{<i}] \sum_{a} x_l^{a} +\frac{\beta}{2N} \sum_{l,l'} \sum_{a,b} x_l^{a} x_l^{b} x_{l'}^{a}x_{l'}^{b} \right]  \right\} \\
& = & e^{ \frac{(N-i) \beta^2}{4N}((N-i)n-n^2) } 
\sum_{\{\underline{x}^{a}\}_{i+1}^N} 
\exp\left\{\beta \left[
\sum_{l} h_l^{\pm}[\mathbf{x}_{<i}] \sum_{a} x_l^{a} +\frac{\beta}{2N} \sum_{a<b} \left( \sum_{l}  x_l^{a} x_l^{b} \right)^2 \right]  \right\}
\end{eqnarray}
\end{widetext}
Using the Hubbard-Stratonovich transformation of the quadratic terms, we can write:

\begin{widetext}
\begin{eqnarray}
\mathbb{E}_{\underline{J}}\left[(\rho_i^{\pm}[\mathbf{x}_{<i}])^n \right] & = & 
c(n,N,i)
\int \prod_{a<b} dQ_{ab} e^{-\frac{N}{2}\beta^2Q_{a,b}^2}
\sum_{\{\underline{x}^{a}\}_{i+1}^N} 
\exp\left\{\beta \left[
\sum_{l} h_l^{\pm}[\mathbf{x}_{<i}] \sum_{a} x_l^{a} +\beta \sum_{a<b} Q_{a,b} \sum_{l}  x_l^{a} x_l^{b} \right]  \right\} \\
& = & 
c(n,N,i)
\int \prod_{a<b} dQ_{ab} e^{-\frac{N}{2}\beta^2Q_{a,b}^2}
\prod_{l} \left[
\sum_{\{\underline{x}^{a}_l\}} 
\exp\left\{\beta \left[
h_l^{\pm}[\mathbf{x}_{<i}] \sum_{a} x_l^{a} +\beta \sum_{a<b} Q_{a,b}  x_l^{a} x_l^{b} \right]  \right\}
\right] \label{eq:before_ansaltz}
\end{eqnarray}
\end{widetext}
where we defined: 

$$c(n,N,i) = e^{ \frac{(N-i) \beta^2}{4N}((N-i)n-n^2) } \left(\frac{2\pi \beta^2}{N}\right)^{\frac{n(n-1)}{2}}.$$ 
The Parisi solutions of the SK model prescribe how to parametrize the matrix of the overlaps $Q$ \cite{10.1142/0271}. The following shows how to obtain neural network architectures based on the replica symmetric (RS) and k-step replica symmetric broken (k-RSB) solutions.

\subsection{Replica Symmetric solution (RS)}
\label{sec:RS}
We assume that the overlaps between the replicas are symmetric under permutations, and the matrix of the overlaps between replicas is parametrized with only one variable $q$:

$$
Q_{a,b}=\begin{cases}
			0, & \text{if $a=b$}\\
            q, & \text{otherwise}
		 \end{cases},
$$
obtaining:

\begin{widetext}
\begin{eqnarray}
\mathbb{E}_{\underline{J}}\left[(\rho_i^{\pm, sym}[\mathbf{x}_{<i}])^n \right] & = & 
c(n,N,i)
\int dq e^{-\frac{n(n-1)N}{4}\beta^2 q^2}
\prod_{l} \left[
\sum_{\{\underline{x}^{a}_l\}} 
\exp\left\{\beta \left[
h_l^{\pm}[\mathbf{x}_{<i}] \sum_{a} x_l^{a} +\beta q \sum_{a<b} x_l^{a} x_l^{b} \right]  \right\} 
\right] \\
& = &
c(n,N,i)
\int dq e^{-\frac{n(n-1)N}{4}\beta^2 q^2}
e^{-\frac{nN\beta^2 q}{2}}
\prod_{l} \left[
\sum_{\{\underline{x}^{a}_l\}} 
e^{\beta \left[
h_l^{\pm}[\mathbf{x}_{<i}] \sum_{a} x_l^{a} + \frac{\beta q}{2} \left(\sum_{a} x_l^{a} \right)^2 \right]} 
\right]\\
& = &
c'(n,N,i)
\int dq e^{-\frac{n(n-1)N}{4}\beta^2 q^2}
e^{-\frac{nN\beta^2 q}{2}}
\prod_{l} \left[\int \frac{dz_l}{\sqrt{2\pi q}} e^{-\frac{z_l^2}{q}}
\sum_{\{\underline{x}^{a}_l\}} 
e^{\beta \left(
h_l^{\pm}[\mathbf{x}_{<i}] +\beta z_l \right) \sum_{a} x_l^{a}} 
\right]\\
& = &
c'(n,N,i)
\int dq e^{-nN\left(\frac{(n-1)}{4}\beta^2 q^2 +\frac{\beta^2 q}{2}\right)}
\prod_{l} \left[\int \frac{dz_l}{\sqrt{2\pi q}} e^{-\frac{z_l^2}{q}}
2^n\cosh^n \left(\beta \left(
h_l^{\pm}[\mathbf{x}_{<i}] +\beta z_l \right)\right) 
\right].\\
\end{eqnarray}
\end{widetext}
Using the limit that $n\rightarrow 0$ we can write:

\begin{widetext}
\begin{eqnarray}
\int \frac{dz_l}{\sqrt{2\pi q}} e^{-\frac{z_l^2}{q}}
2^n\cosh^n \left(\beta \left(
h_l^{\pm}[\mathbf{x}_{<i}] +\beta z_l \right)\right) = e^{n \int \frac{dz_l}{\sqrt{2\pi q}} e^{-\frac{z_l^2}{q}}
\log 2\cosh \left(\beta \left(
h_l^{\pm}[\mathbf{x}_{<i}] +\beta z_l \right)\right)}.
\label{eq:gauss_n0}
\end{eqnarray}
\end{widetext}
obtaining:

\begin{widetext}
\begin{eqnarray}
\log (\rho_i^{\pm, sym}[\mathbf{x}_{<i}]) & = & 
\lim_{n\rightarrow 0} \frac{1}{n} \log \left( c'(n,N,i)
\int dq e^{-\frac{n(n-1)N}{4}\beta^2 q^2}
e^{-\frac{nN\beta^2 q}{2}}
e^{n \sum_l 
\int \frac{dz_l}{\sqrt{2\pi q}} e^{-\frac{z_l^2}{q}}
\log 2\cosh \left(\beta \left(
h_l^{\pm}[\mathbf{x}_{<i}] +\beta z_l \right)\right)
} 
\right)\\
& = &
\log(c''(N,i)) + 
\left( +\frac{N}{4}\beta^2 q^2_0 
-\frac{N\beta^2 q_0}{2}
+ \sum_l 
\int \frac{dz_l}{\sqrt{2\pi q_0}} e^{-\frac{z_l^2}{q_0}}
\log 2\cosh \left(\beta \left(
h_l^{\pm}[\mathbf{x}_{<i}] +\beta z_l \right)\right)
\right) \\
& \doteq &  
c(N,i, q_0) -
\sum_l 
\int \frac{dz_l}{\sqrt{2\pi q_0}} e^{-\frac{z_l^2}{q_0}}
\log \sigma \left(\beta \left(
2h_l^{\pm}[\mathbf{x}_{<i}] +2\beta z_l \right)\right)
\end{eqnarray}
\end{widetext}

In the second line, I use the saddle point methods to evaluate the integral over $q$, assuming that the single maximum value $q_0$ does not depend on the input values $\mathbf{x}_{<i}$ in the set of $h_l^{\pm}[\mathbf{x}_{<i}]$. It is a bold assumption to be verified {\it a posteriori} on the quality of the neural network architectures performances. 
In the third line, we use the identity $\log\cosh(x) = 2x - \log\sigma(2x)$ and the elements that are equals between $\log(\rho^+)$ and $\log(\rho^-)$ are simplified. We introduced the $\log\sigma$ non-linear operator for computational reasons.

Now we consider the following approximation of the Gaussian convolution:

\[
\int dz e^{-z^2}
\log \sigma \left(\beta \left(
h +\beta \sqrt{q}z \right)\right) \sim b_0 + w_0*\log \sigma(b_1 + w_1 h), 
\]

where $(b_0, w_0, b_1,w_1)$ are free parameters to be determined. In the sec.\ref{sec:num_evidence} of the SI a numerical analysis of the quality of this approximation is shown. 
Putting together all the pieces, we can parameterize the conditional probability as:

\begin{multline}
Q^{RS}\left(x_{i}=1|\mathbf{x}_{<i}\right) = \sigma\left( 
    x_i^1(\mathbf{x}_{<i}) +\log(\rho_i^+) - \log(\rho_i^-)
\right) \\
 = \sigma \bigg( x_i^1(\mathbf{x}_{<i}) + \sum_{l^+=i+1}^{N}  w_0^{i,l^+} \log\sigma(b_1^{i,l^+} +
 w_1^{i,l^+} x_{i,l^+}^1(\mathbf{x}_{<i}))+ 
 + \sum_{l^-=i+1}^{N}  w_0^{i,l^-} \log\sigma(b_1^{i,l^-} + w_1^{i,l^-} x_{i,l^-}^1(\mathbf{x}_{<i})
 \bigg) 
\end{multline}

where the set of $(\mathbf{b},\mathbf{w})$ are free variational parameters to learn in the training process. 

\subsection{K-step Replica Symmetric Breaking (k-RSB) solution}
\label{sec:rsb}
Assuming that the replica symmetry is broken, we can use the ansatz called 1-step replica symmetric breaking (1RSB), where the overlaps between replicas are divided into $m$ blocks:

\begin{eqnarray}
    Q_{a,b}=\begin{cases}
			q_1, & \text{if } I(a/m)=I(b/m) \\
            q_0. & \text{if } I(a/m) \neq I(b/m).
		 \end{cases}
\end{eqnarray}
With the above ansatz, we can compute the following quantities:

\begin{align}
\begin{split}
    \sum_{ab} Q_{ab} x_{a} x_{b}  &  = \frac{1}{2} \bigg[ q_0 \left( \sum_{a}x_a\right)^2 +  (q_1-q_0) \sum_{\text{blocks}}  \left( \sum_{a \in \text{block}}x_a\right)^2   - nq_1\bigg] 
\end{split}
\\
\sum_{ab} Q_{ab}^2 & =  n^2 q_0^2 + nm(q_1^2 - q_0^2) -n q_1^2.
\end{align}

The equation \ref{eq:before_ansaltz} becomes:

\begin{align*}
& \mathbb{E}_{\underline{J}} \left[(\rho_i^{\pm, 1RSB})^n \right] =  \\[1ex]
\begin{split}
& = c(n,N,i) \int dq_1 dq_0 e^{\frac{N}{2}\beta^2 \left[n^2 q_0^2 + nm(q_1^2 - q_0^2) -n q_1^2 \right]} 
\prod_{l} \bigg[ \sum_{\{\underline{x}^{a}_l\}} e^{ \beta \big[ h_l^{\pm} \sum_{a} x_l^{a} +\beta q_0 \left( \sum_{a} x_p^{a} \right)^2 + \beta (q_1-q_0) \sum_{\text{blocks}} \left( \sum_{a \in \text{block}}x_l^{a}\right)^2  -n q_1 \bigl]}  \bigg] 
\end{split}\\ 
\begin{split}
& = c(n,N,i) \int dq_1 dq_0 e^{\frac{N}{2}\beta ^ 2 \left[n^2 q_0^2 + nm(q_1^2 - q_0^2) -n q_1^2 -n q_1\right]} 
\prod_{l} \bigg[ \sum_{\{\underline{x}^{a}_l\}} \int dP_{z_l} \prod_{k=1}^{n/m} \int dP{y_{lk}}  e^{\beta \big[h_l^{\pm} \sum_{a} x_l^{a} + \beta z_l \sum_{a}x_l^{a} + \beta \sum_{\text{blocks}}  y_{lk} \sum_{a \in \text{block}}x_l^{a}\bigl]}  \bigg] 
\end{split}\\ 
\begin{split}
& = c(n,N,i) \int dq_1 dq_0 e^{\frac{N}{2}\beta^2 \left[n^2 q_0^2 + nm(q_1^2 - q_0^2) -n q_1^2 -n q_1\right]} 
\prod_{l} \bigg[ \int dP_{z_l}  \prod_{k=1}^{n/m} \int dP_{y_{lk}} \cosh^m\bigg(\beta \big[h_l^{\pm}+ \beta z_l +\beta y_{lk}\bigl]  \bigg)  \bigg]
\end{split}\\ 
\begin{split}
& = c'(n,N,i) + c(n,N,i) \int dq_1 dq_0 e^{\frac{N}{2}\beta\left[n^2 q_0^2 + nm(q_1^2 - q_0^2) -n q_1^2 -n q_1\right]} 
\prod_{p} \int dP_{z_l}  \exp \bigg\{ \frac{n}{m} \log \bigg( \int dP_{y_{l}} \cosh^m\bigg(\beta \big[h_l^{\pm}+ \beta z_l + \beta y_{l}\bigl]  \bigg)  \bigg) \bigg\},
\end{split}\\ 
\end{align*}
where we defined:

\begin{align}
    dP_{z_l} & = \frac{dz_l}{\sqrt{2\pi q_0}}e^{\frac{z^2}{2q_0}}\\
    dP_{y_{l}} & = \frac{dy_{l}}{\sqrt{2\pi (q_1-q_0)}}e^{\frac{y_{l}^2}{2 (q_1-q_0)}}.
\end{align}
Considering $N \gg 1$ and $n\rightarrow 0$ to use the saddle point methods and the identity in Eq.~(\ref{eq:gauss_n0}), we can write:

\begin{widetext}
\begin{align}
\log (\rho_i^{\pm, 1RSB}) & = 
\lim_{n\rightarrow 0} \frac{1}{n} \log \left(\mathbb{E}_{\underline{J}} \left[(\rho_i^{\pm, 1RSB})^n \right]  \right) \\
& = c_i +  \text{Extr}_{q_0, q_1} \bigg[ c'_i(N,n,q_0, q_1) 
+ \frac{1}{m} \sum_{l} \int dP_{z_l} \log \bigg( \int dP_{y_{l}} \cosh^m\bigg(\beta \big[h_l^{\pm}+ \beta z_l + \beta y_{l}\big]  \bigg)  \bigg)
\bigg].
\end{align}
\end{widetext}

The above integrals are rewritten as the following:

\begin{align}
& \int dP_{z_l} \log \bigg( \int dP_{y_{l}}  \cosh^m\bigg(\beta \big[h_l^{\pm}+\beta z_l + \beta  y_{l}\big]  \bigg)  \bigg) 
 = \\
& \int dP_{z_l} \log \biggl( \int dP_{y_{l}} e^{ m \log \cosh \left(\beta \left[h_l^{\pm}+ \beta  z_l + \beta  y_{l}\right]  \right) } \biggr) 
 = \\
& \beta h_{l}^{\pm} + \int dP_{z_l} \log \biggl( \int dP_{y_{l}} e^{\beta^2 m y_{l} - m \log \sigma \left(\beta \left[h_l^{\pm}+ \beta z_l +\beta y_{l}\right]  \right) } \biggr) 
\end{align}
We have two nested Gaussian convolutions.

The integrals over the variables $(z_l, y_l)$ are approximated in the same way as the approach used previously for the RS case: the number of free parameters increases to have feed-forward functions without integrals. First, we consider the following maps:

\begin{align}
    & \log \int dP_{y_{l}} e^{ \beta^2 m y_{l} - m \log \sigma \left(\beta \left[h_l^{\pm}+\beta z_l +\beta y_{l}\right]  \right) }  \approx\\
    & \log \bigg( \hat{b}_0^{l^{\pm}} + \hat{w}_0^{l^{\pm}} e^{\hat{b}_1^{l^{\pm}} + \hat{w}_1^{l^{\pm}} \log \sigma (\hat{b}_2^{l^{\pm}} + \hat{w}_2^{l^{\pm}} (h_l^{\pm}+ \beta  z_l))} \bigg) = \\
    & \log(b_0^{l^{\pm}}) + \log(1 + e^{ \log( \frac{\hat{w}_0^{l^{\pm}}}{b_0^{l^{\pm}}} ) + \hat{b}_1^{l^{\pm}} + \hat{w}_1^{l^{\pm}} \log \sigma (\hat{b}_2^{l^{\pm}} + \hat{w}_2^{l^{\pm}} (h_l^{\pm}+ \beta  z_l))})=\\
    & \log(\hat{b}_0^{l^{\pm}}) - \log\sigma( \log( \frac{\hat{w}_0^{l^{\pm}}}{b_0^{l^{\pm}}} ) + \hat{b}_1^{l^{\pm}} + \hat{w}_1^{l^{\pm}} \log \sigma (\hat{b}_2^{l^{\pm}} + \hat{w}_2^{l^{\pm}} (h_l^{\pm}+ \beta  z_l)))=\\
    & b_0^{l^{\pm}} + w_0^{l^{\pm}} \log\sigma(b_1^{l^{\pm}} + w_1^{l^{\pm}} \log \sigma (b_2^{l^{\pm}} + w_2^{l^{\pm}} (h_l^{\pm}+ \beta  z_l)))
\end{align}
Where in the last equation I renamed the variational parameters, for instance, $ b_0^{l^{\pm}} = \log(\hat{b}_0^{l^{\pm}})$, $w_0^{l^{\pm}} =-1$, etc. Then once again:

\begin{align}
        & \hat{b}_0^{l^{\pm}} + \hat{w}_0^{l^{\pm}} \int dP_{z_l} \log \sigma \bigg(\hat{b}_1^{l^{\pm}} + \hat{w}_1^{l^{\pm}} \log \sigma (\hat{b}_2^{l^{\pm}} + \hat{w}_2^{l^{\pm}} (h_l^{\pm}+ \beta  z_l)) \bigg) \approx \\
        & b_0^{l^{\pm}} + w_0^{l^{\pm}} \log \sigma (b_1^{l^{\pm}} + w_1^{l^{\pm}} \log \sigma (b_2^{l^{\pm}} + w_2^{l^{\pm}} (h_l^{\pm}))).
\end{align}
The set of parameters $(\mathbf{b_0^{{\pm}}},\mathbf{w_0^{{\pm}}},\mathbf{b_1^{{\pm}}},\mathbf{w_1^{{\pm}}},\mathbf{b_2^{{\pm}}},\mathbf{w_2^{{\pm}}})$ are free parameters to be determined by the learning procedures. The Gaussian convolution integrals are substituted by feed-forward non-linear operations, enlarging the space of parameters. In section \ref{sec:num_evidence} numerical evidence for the goodness of the above approximations is shown. For the 1RSB case, we use the following architecture:

\begin{multline*}
    Q^{1RSB}\left(x_{i}=1|\mathbf{x}_{<i}\right) = \sigma\left( 
        x_i^1(\mathbf{x}_{<i}) +\log(\rho_i^+) - \log(\rho_i^-)
    \right) 
     = \sigma \bigg( x_i^1(\mathbf{x}_{<i}) + \\ \sum_{l^+=i+1}^{N}  w_0^{i,l^+} \log\sigma(b_1^{i,l^+} + 
     w_1^{i,l^+} \log\sigma(b_2^{i,l^+} +
     w_2^{i,l^+}  x_{i,l^+}^1(\mathbf{x}_{<i})))+ \sum_{l^-=i+1}^{N}  w_0^{i,l^-} \log\sigma(b_1^{i,l^-} + w_1^{i,l^-} \log\sigma(b_2^{i,l^+} +
     w_2^{i,l^+} x_{i,l^-}^1(\mathbf{x}_{<i})))
     \bigg) 
\end{multline*}   
The generalization of the $\rho^{\pm}$ parametrization to the k-RSB case is straightforward. For instance, for 2-RSB we have:

\begin{equation*}
    \log \rho^{\pm, 2RSB} \left(x_{i}=1|\mathbf{x}_{<i}\right)  =  
   \sum_{l^{\pm}} w_0^{i,l^{\pm}} \log\sigma(b_1^{i,l^{\pm}} +
   w_1^{i,l^{\pm}} \log\sigma(b_2^{i,l^{\pm}} +
    w_2^{i,l^{\pm}}( \log\sigma(b_3^{i,l^{\pm}} +
    w_3^{i,l^{\pm}}x_{i,l^{\pm}}^1(\mathbf{x}_{<i})))))
\end{equation*}

\subsection{Numerical evidence of integrals approximations}
\label{sec:num_evidence}
\begin{figure}[h]
    \centering
    \includegraphics[width=1\textwidth]{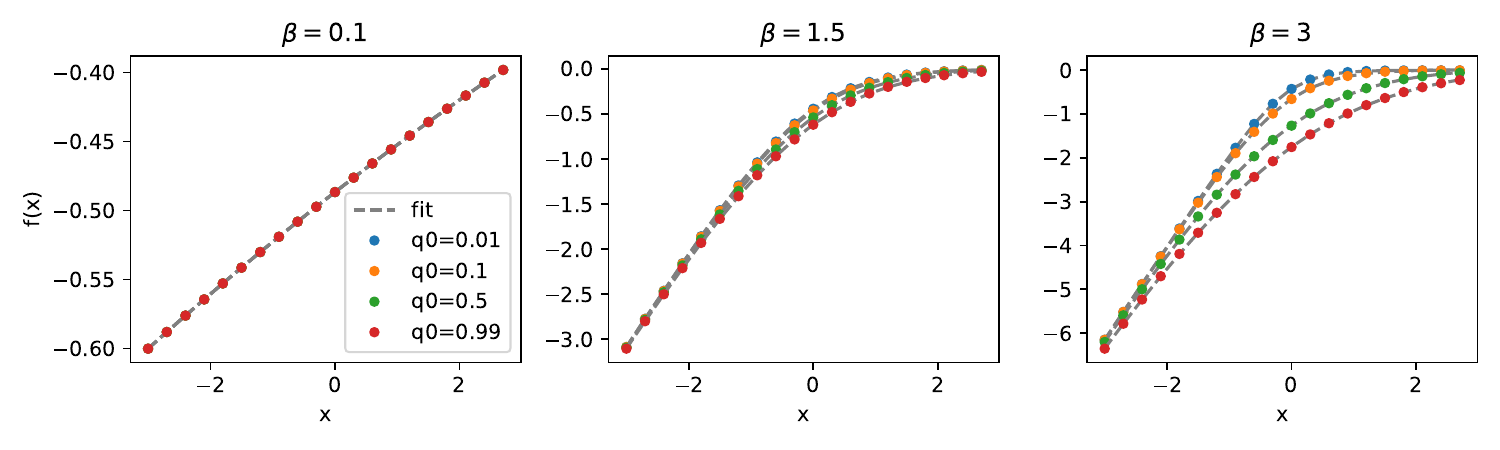}
    \caption{Fit of Eq.~(\ref{eq:rs_approx}) at different values of $\beta$ and $q_0$. The free parameters are $b_0, w_0, b_1, w_1$. The points are numerical evaluations of the integrals, and the grey lines are the corresponding fit using the approximated functions. The results show the robustness of the approximation used. }
    \label{fig:rs_approx}
\end{figure}

\begin{figure}[h]
    \centering
    \includegraphics[width=1\textwidth]{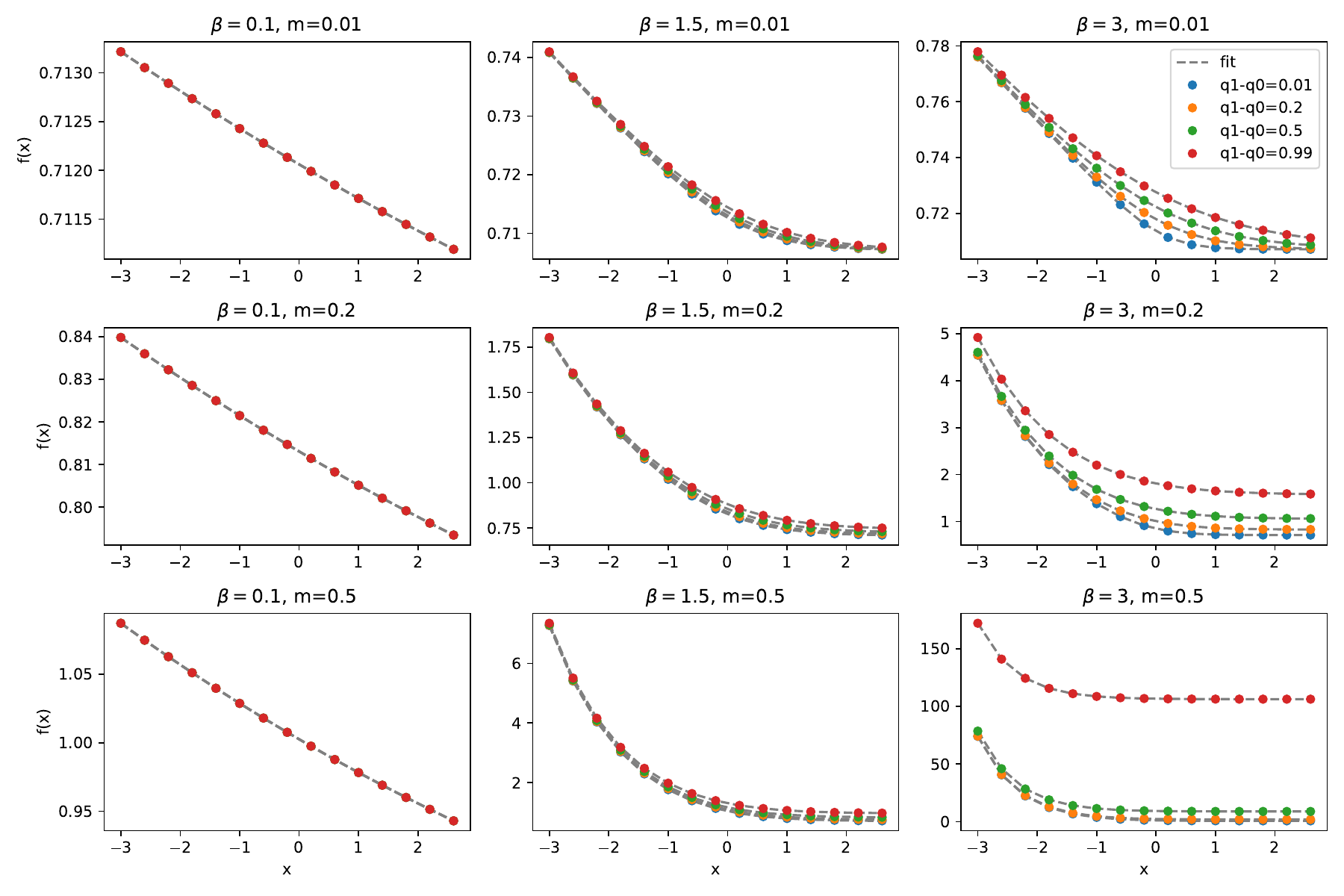}
    \caption{Fit of Eq.~(\ref{eq:1rsb_approx}) at different values of $beta$ $m$, and $q_1 - q_0$. The free parameters are $b_0, w_0, b_1, w_1, b_2, w_2$.  The points are numerical evaluations of the integrals, and the gray lines are the corresponding fit using the approximated functions.  The results show the robustness of the approximation used. }
    \label{fig:1rsb_approx}
\end{figure}

In this section, I show the numerical evidence of some of the approximations used to represent the non-linear operations derived for the SK model. In the RS case, see Subsec. \ref{sec:RS}, the non-linear operation is approximated as in the following:

\begin{equation}
    \label{eq:rs_approx}
    f(x) = \int \frac{dz}{\sqrt{2 \pi q_0}} e^{-\frac{z^2}{q_0}}
\log \sigma \left(\beta \left(
x +\beta \sqrt{q}z \right)\right) \sim b_0 + w_0*\log \sigma(b_1 + w_1 h). 
\end{equation}
The results in Fig.~\ref{fig:rs_approx} of the fit, using the non-linear least squares method, of the $f(x)$ with the r.h.s of the above expression, with free parameters $b_0, w_0, b_1, w_1$, are shown. The fits, for different values of $\beta$ and $q_0$, show the robustness of the approximation used.  In the 1RSB case, see subsection \ref{sec:rsb}, one of the non-linear operations is approximated as in the following:

\begin{equation}
    f(x) = \log \int  \frac{dz}{\sqrt{2 \pi (q_1-q_0)}} e^{-\frac{z^2}{(q_1-q_0)}} e^{ \beta^2 m z - m \log \sigma \left(\beta x+\beta^2 z \right) }  \approx b_0  + w_0  \log\sigma(b_1  + w_1  \log \sigma (b_2  + w_2  x))
    \label{eq:1rsb_approx}
\end{equation}
As before, the function $f(x)$ is fitted with the r.h.s of the above expression, and the results in Fig.~\ref{fig:1rsb_approx} show the robustness of the approximation used for different values of $(q_1-q_0)$, $m$, and $\beta$.

\section{Monte Carlo sampling of the SK model}
In the main text, in the section results, in order to check the correlation between the weights of the first layer of the SK$_{1RSB}$ architecture with the coupling of the Hamiltonian, 10,000 samples were generated through the Monte-Carlo Metropolis algorithm \cite{doi:10.1063/1.1887186}. The samples are generated for a system with $N=200$ variables at $\beta=2$. Each configuration is sampled after $200*N$ spin trail flips.  I observe that, for the system under study, for this number of trail spin flips of the Metropolis algorithm the autocorrelation time between the configurations drops below 0.5. The training of the SK$_{1RSB}$ was done minimizing, over the parameters of the SK$_{1RSB}$, the average of the negative log-likelihood (LL) of the generated $m$ samples:
 
\begin{equation}
    LL = - \frac{1}{m} \sum_{m}\log(Q_{1RSB})
\end{equation}  
\bibliography{main}